\documentclass[11pt,a4paper]{article}

\usepackage[T1]{fontenc}
\usepackage{amssymb,amsmath,amsthm}
\usepackage{graphicx,color}
\usepackage{a4wide}
\usepackage[authoryear,round]{natbib}

\theoremstyle{definition}
\newtheorem{alg}{Algorithm}

\begin{document}
\title{\bf On the estimation of parameters of a spheroid distribution from planar sections}
\author{Markus Baaske$^{1,2}$, Felix Ballani$^1$, and Alexandra Illgen$^3$\\[5mm]
	${}^1$Institute of Stochastics, Technische Universit{\"a}t Bergakademie Freiberg,\\ 
	D-09596 Freiberg, Germany\\[5mm]
	${}^2$Michael Stifel Center Jena, Ernst-Abbe-Platz 2--4, D-07743 Jena, Germany\\[5mm]
	${}^3$Institute of Materials Engineering, Technische Universit{\"a}t Bergakademie Freiberg,\\ 
	D-09596 Freiberg, Germany\\[5mm]}

\maketitle

\begin{abstract}
We study two different methods for inferring the parameters of a spheroid distribution from planar sections of a stationary spatial system of spheroids: one method first unfolds non-parametrically the joint size-shape-orientation distribution of the observable ellipses in the plane into the joint size-shape-orientation distribution of the spheroids followed by a maximum likelihood estimation of the parameters; the second method directly estimates these parameters based on statistics of the observable ellipses using a quasi-likelihood approach. As an application we consider a metal-matrix composite with ceramic particles as reinforcing inclusions, model the inclusions as prolate spheroids and estimate the parameters of their distribution from planar sections.
\end{abstract}



\section{Introduction}\label{sec:intro}
Although nowadays various imaging techniques like X-ray computed tomography are available by which three-dimensional specimens, e.\,g. composite materials, can be represented truly three-dimensional for further analysis this is sometimes too expensive or not applicable due to the type of material, or, the resolution of the sampling technique is too low in order to distinguish very small substructures or inclusions. However, the geometry of such spatial structures still allows for investigation via planar sections where then often more highly resolving sampling techniques are available \citep{ref:Nagel2010}. The methodology aimed to draw inference from information contained in planar sections about the spatial geometrical structure is part of stereology \citep{ref:OhserMuecklich2000,ref:Nagel2010,ref:CSKM2013}. 

\begin{figure}
\begin{center}
\includegraphics[width=0.6\textwidth]{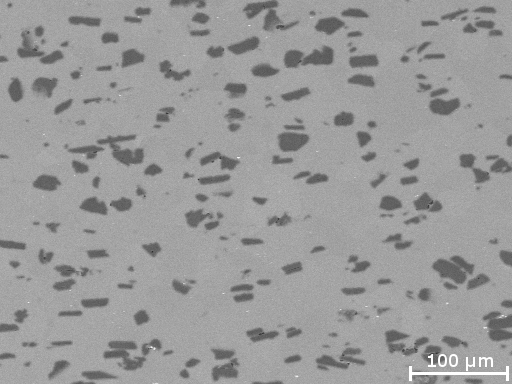}
\end{center}
\caption{Planar section (cutout) of a metal-matrix composite with reinforcing ceramic particles resulting from scanning electron microscopy.}
\label{fig:data}
\end{figure}

We were faced with this kind of situation when investigating and modelling metal-matrix composites with ceramic particles as reinforcing inclusions (see Figure \ref{fig:data}) with respect to their fatigue behaviour. The kind of model intended for the fatigue behaviour \citep{ref:BaaskeEtAl2018Book} is based on a model for the spatial configuration of the ceramic inclusions.
Besides a model for the spatial arrangement of the inclusions \citep[based on the force-biased algorithm for ellipsoids, see][]{ref:BezrukovStoyan2006} this requires the availability of a reasonable model for the distribution of the inclusions in the sense of a grain distribution \citep[][Sect. 6.5.1]{ref:CSKM2013}. Under the assumption of spatial stationarity it is justified to speak of the \emph{typical} inclusion, i.\,e., the distribution of which does not depend on the particular (stationary) spatial arrangement of the inclusions \citep[cf., e.\,g.,][Sect. 4.2]{ref:SchneiderWeil2008}; likewise this applies for the distribution of the typical section profile of those inclusions which are intersected by some plane. With this reasoning in mind and since in the present paper we solely focus on the distribution of the typical inclusion we will throughout the paper work with a stationary \emph{Poisson} particle process \citep[][Sect. 4.1]{ref:SchneiderWeil2008} for the spatial arrangement even when processing the data (where the Poisson assumption is certainly not true), which is an established approach in stereology \citep[][p. 437]{ref:CSKM2013}. 

In particular, due to the possible flexibility w.r.t. shape, we decided to model the typical inclusion as a random \emph{ellipsoid} with randomness in size, shape and orientation. Thus it is then natural to model the intersected inclusions in the observable planar sections as \emph{ellipses} for which size, shape and orientation (in the plane) can be estimated from digital images.  

Dating back to \citet{ref:Wicksell1926} the respective stereological objective is to infer the joint distribution of the ellipsoids' size and shape (and also orientation) from the ellipses observable in the planar section which turned out to be solvable completely and uniquely only in case the ellipsoids are all either prolate or oblate spheroids (ellipsoids of revolution), see the pioneering work of \citet{ref:CruzOrive1976}. The general solution of unfolding the joint size-shape-orientation distribution of the observable ellipses in the plane into the joint size-shape-orientation distribution of the spheroids was given by \citet{ref:BenesEtAl1997} and \citet{ref:BenesKrejcir1997}, see also \citet[Ch. 6]{ref:BenesRataj2004}. Hence, from the empirical 3D joint size-shape-orientation distribution, typically available as a histogram, one may estimate the parameters of a parametric spheroid distribution, for instance by the maximum likelihood method.

A different way is to estimate the parameters based on statistics directly available from the 2D joint size-shape-orientation distribution of the ellipses. A general approach for such a kind of estimation would be to find those model parameters which lead to statistics most similar to the observed ones in the sense of a least squares or minimum contrast approach. Instead of minimizing a distance a related alternative is to find the root of an estimation equation. A particular such estimation equation is based on the so-called quasi-score function which leads to quasi-likelihood estimation, see \citet{ref:Heyde1997} for the general theory. A general problem which holds both for finding a minimum or a root is to explore efficiently the space of possible parameters in case the objective function can be determined only by simulations. For quasi-likelihood estimation an approach is given in \citet{ref:BaaskeEtAl2014}.

The aim of the present paper, motivated by the following facts, is to compare the (possibly more established) way of estimating the parameters by the maximum likelihood method based on the stereologically unfolded joint size-shape-orientation distribution with the quasi-likelihood estimation approach. On the one hand, quasi-likelihood estimation based on simulations is much more involved than unfolding followed by maximum likelihood estimation. On the other hand, unfolding as the solution of an ill-posed inverse problem has a tendency to corrupt the subsequent parameter estimation whereas in quasi-likelihood estimation the available information contained in the employed statistics is used in an optimal way, as demonstrated in \citet{ref:BaaskeEtAl2014} with an example in a spatial context, leading to more precise estimation results. 

The paper is organized as follows. In Section \ref{sec:model} we introduce the parametric model for a spheroid distribution for which we aim to estimate parameters. Furthermore, in Section \ref{sec:unfold+ML}, we first summarize the necessary facts for the trivariate unfolding related to prolate spheroids and give then details for the subsequent maximum likelihood estimation. In Section \ref{sec:QL} we sketch the ideas of the quasi-likelihood estimation approach. Then, in Section \ref{sec:sim}, we compare both approaches by means of a simulation study. Finally, in Section \ref{sec:appl} we apply them to a planar section of a system of particles and end up with some conclusions in Section \ref{sec:concl}.
%
\section{A parametric spheroid distribution}\label{sec:model}
In view of the two basic possibilities of taking the ellipsoids for a reasonable stereological unfolding all either as prolate or as oblate spheroids and respecting findings on the reinforcing particles in a metal-matrix composite \citep{ref:BorbelyEtAl2004} quite similar to that in our application in Section \ref{sec:appl} we restrict henceforth to \emph{prolate} spheroids with lengths $a\geq b=c$ of the semi-axes. The shape factor (aspect ratio) $s$ of such a prolate spheroid is then defined as $s=c/a$ and satisfies $0<s\leq1$. Finally, the orientation of a prolate spheroid as the direction of the axis of revolution can be described in terms of spherical coordinates $(\vartheta,\varphi)$ w.r.t. a fixed axis $u$ with polar angle $\vartheta\in[0,\pi/2)$ and azimuthal angle $\varphi\in[0,2\pi)$. Hence, besides location, a (prolate) spheroid is described for instance by $(a,s,\vartheta,\varphi)$ or $(c,s,\vartheta,\varphi)$, respectively. 

To simplify matters we assume that the distribution of the orientations is independent of that of the sizes and shapes. Furthermore, the production process \citep{ref:MuellerEtAl2015} of the metal-matrix composite justifies the assumption that the distribution of the system of spheroids is invariant w.r.t. rotations about the fixed axis $u$. One such model for the random orientation of a spheroid respecting this kind of invariance is the `Schladitz distribution' \citep{ref:FrankeEtAl2016} which has probability density function (p.d.f.)
\begin{equation}\label{eqn:pbeta}
h_{\beta}(\vartheta,\varphi)=\frac{1}{2\pi}\cdot\frac{1}{2}\frac{\beta\sin\vartheta}{(1+(\beta^2-1)\cos^2\vartheta)^{\frac{3}{2}}},\quad\vartheta\in[0,\pi),\varphi\in[0,2\pi),
\end{equation}
see also \citet[Eq. (7.11)]{ref:OhserSchladitz2009}. In this model $\beta>0$ is an anisotropy parameter in the sense that for decreasing $\beta<1$ the spheroids tend to be more and more parallel to the $u$-axis, and, respectively, for increasing $\beta>1$ the spheroids tend to be more and more parallel to the plane perpendicular to $u$, where $\beta=1$ is the case of isotropically distributed directions.

Since the population of ceramic particles suggests a possible dependence of size and shape of the corresponding representing ellipsoids and a typical size distribution for granular inclusions in materials science is the log-normal distribution, in particular, \citet{ref:BorbelyEtAl2004} figured out that the log-normal distribution is a good model for the particle sizes in a composite material similar to that considered in Section \ref{sec:appl}, we aim to work with the following model for the length $a$ of the semi-major axis and the shape factor $s$. Let $(\xi,\eta)$ be a bivariate normally distributed random vector with mean vector $\mu=(\mu_1,\mu_2)\in\mathbb{R}^2$ and variance-covariance-matrix
\begin{equation}\label{eqn:Sigma}
\Sigma=\begin{pmatrix}\sigma_1^2&\varrho\sigma_1\sigma_2\\ \varrho\sigma_1\sigma_2&\sigma_2^2\end{pmatrix},\quad\sigma_1,\sigma_2\geq0,-1\leq\varrho\leq1.
\end{equation} 
Then let
\begin{equation}\label{eqn:as}
a=\exp(\xi),\quad s=\frac{1}{1+\exp(-\eta)}.
\end{equation}
The parameter $\varrho$ accounts for a possible dependence of size and shape, in particular, $a$ and $s$ are independent if and only if $\varrho=0$. Furthermore, for $\sigma_1=0$ or $\sigma_2=0$ the cases of a deterministic size or a deterministic shape are included. Note that the approach (\ref{eqn:as}) ensures that the shape factor $s$ is always between $0$ and $1$ and, implying $a>c$, that the corresponding spheroids are always prolate. 

All in all we end up with a parametric model for the size-shape-orientation distribution of a random spheroid including the six parameters $(\mu_1,\mu_2,\sigma_1,\sigma_2,\varrho,\beta)$. If a system of $n$ spheroids (for instance associated with the ceramic inclusions) was observable directly in terms of $(a_l,s_l,\vartheta_l)$, $l=1,\ldots,n$, then the usual way of parameter estimation would be the maximum likelihood method since the likelihood is available. In what follows we discuss two ways how the model parameters might be estimated in case only planar sections of the inclusions are given.
%
\section{Maximum likelihood estimation after trivariate unfolding}\label{sec:unfold+ML}
\subsection{Unfolding}\label{sec:unfold}
Let $\Psi_V$ be a stationary spheroid process with intensity $\lambda_V$ which
is intersected by a random plane $\mathcal{H}$ with normal direction $v$ perpendicular to the reference direction $u$, i.\,e. $v$ has polar angle $\pi/2$ and uniformly on $[0,2\pi)$ distributed azimuthal angle ($\overline{\varphi}$) w.r.t. $u$ \citep[termed `vertical uniform random section' in case $u=(0,0,1)$, see][]{ref:BenesRataj2004}. 
\begin{figure}
\centering
\includegraphics[width=0.4\textwidth]{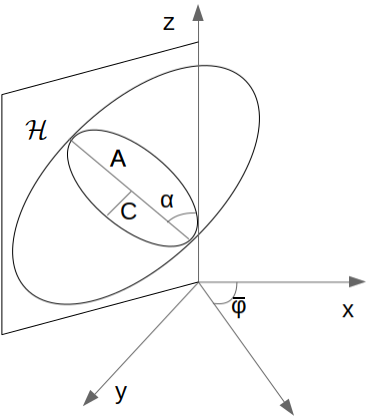}
\caption{Section ellipse of a spheroid within a vertical uniform random section plane.}
\label{fig:vertSect}
\end{figure}
In case a spheroid is hit by $\mathcal{H}$ the intersection of both is an ellipse with lengths $A$ and $C$, $A\geq C$, of the two semi-axes, related shape factor $S=C/A$ and angle $\alpha$ between the semi-major axis and the reference direction $u$, see Figure \ref{fig:vertSect} for an illustration. The cumulative distribution function (c.d.f.) $G(C,S,\alpha)$ of the triple $(C,S,\alpha)$ related to the typical intersection ellipse and the c.d.f. $H(c,s,\vartheta)$ of the triple $(c,s,\vartheta)$ related to the typical spheroid are related to each other by the integral equation
\begin{equation}\label{eqn:integraleq}
\lambda_AG(C,S,\alpha)=\frac{4}{\pi}\lambda_V\int\mathbf{1}_{[C,\infty)}(c)(c-\sqrt{c^2-C^2})K_0(\alpha,S,\vartheta,s)\,\mathrm{d}H(c,s,\vartheta),
\end{equation}
see \citet[Thm. 6.17]{ref:BenesRataj2004}, where $\lambda_A$ is the intensity of the stationary intersection ellipse process and $K_0(\alpha,S,\vartheta,s)$ is some function, the explicit form of which is omitted here and can be found in the cited theorem. Based on the observations $(C_l,S_l,\alpha_l)$, $l=1,\ldots,n$, the objective of unfolding is to reconstruct $H(c,s,\vartheta)$ by solving (\ref{eqn:integraleq}) for $H(c,s,\vartheta)$. The numerical solution presented in \citet[Sect. 6.3.3]{ref:BenesRataj2004}, which we aim to apply in what follows, transforms the integral equation (\ref{eqn:integraleq}) into a system of linear equations by discretization and solves this system with the help of the expectation-maximization algorithm. The discretization is based on classes
\begin{equation*}
D_{ijk}=\{(c,s,\vartheta):\,c_{i-1}<c\leq c_i,s_{j-1}<s\leq s_j,\vartheta_{k-1}<\vartheta\leq\vartheta_k\},
\end{equation*}
$i=1,\ldots,N_c$, $j=1,\ldots,N_s$, $k=1,\ldots,N_{\vartheta}$, and, respectively,
\begin{equation*}
\tilde{D}_{IJK}=\{(C,S,\alpha):\,C_{I-1}<C\leq C_I,S_{J-1}<S\leq S_J,\alpha_{K-1}<\alpha\leq\alpha_K\},
\end{equation*}
$I=1,\ldots,N_C$, $J=1,\ldots,N_S$, $K=1,\ldots,N_{\alpha}$, where for simplicity we work with $N_C=N_c$, $N_S=N_s$, $N_{\alpha}=N_{\vartheta}$ and $\tilde{D}_{ijk}=D_{ijk}$ for all $(i,j,k)$.

From the planar observations first a normalized trivariate histogram with relative frequencies $g_{ijk}$ related to class $D_{ijk}$ is determined. Then unfolding results in another normalized trivariate histogram with relative frequencies $h_{ijk}$ which is a kind of kernel density estimation of the probability density function of $H(c,s,\vartheta)$.

An implementation of the unfolding procedure as in \citet[Sect. 6.3.3]{ref:BenesRataj2004} adapted to the case of prolate spheroids is available within the contributed \textsf{R} package \textsf{unfoldr}, see \citet{ref:Baaske2017}. 
%
\subsection{Maximum likelihood estimation}\label{sec:ML}
Since the result of the unfolding is the trivariate size-shape-orientation histogram $(h_{ijk})$ a subsequent maximum likelihood estimation has to be based on that and not on the original likelihood related to $(c,s,\vartheta)$ or, equivalently, $(a,s,\vartheta)$.

In the model introduced in Section \ref{sec:model} we assume that the orientation of the spheroids is independent of size and shape. Hence the parameter $\beta$ of the orientation distribution can be estimated separately. The respective log-likelihood reads
\begin{equation*}
\log\left(L\left((h_{\cdot\cdot k});\beta\right)\right)=\sum_{k=1}^{N_{\vartheta}}h_{\cdot\cdot k}\log\bigg(\mathsf{P}_{\!\beta}\big((\vartheta_{k-1},\vartheta_k]\big)\bigg)
\end{equation*}
where $\mathsf{P}_{\!\beta}$ denotes the probability measure related to the p.d.f. $h_{\beta}$, 
\[
h_{\cdot\cdot k}=\sum_{i=1}^{N_c}\sum_{j=1}^{N_s}h_{ijk}
\]
is the marginal relative frequency in bin $(\vartheta_{k-1},\vartheta_k]$, $k=1,\ldots,N_{\vartheta}$, and $N_c$, $N_s$ and $N_{\vartheta}$ are the numbers of bins for, respectively, $c$, $s$ and $\vartheta$ as in Section \ref{sec:unfold}.

Furthermore, with $f_{\mu,\Sigma}$
the p.d.f. of a bivariate normal distribution with mean $\mu=(\mu_1,\mu_2)$ and variance-covariance matrix $\Sigma$ as in (\ref{eqn:Sigma}), the joint p.d.f. of the length $c$ of the semi-minor axis and the shape factor $s$ is
\begin{equation*}
h_{\mu,\Sigma}(c,s)=\frac{1}{c\,s\,(1-s)}\,f_{\mu,\Sigma}(\log(c/s),\log(s/(1-s)))
\end{equation*}
since the reverse transform of $(\xi,\eta)\mapsto(c,s)=(\exp(\xi)/(1+\exp(-\eta)),1/(1+\exp(-\eta))$ as in (\ref{eqn:as}) is $(c,s)\mapsto(\xi,\eta)=(\log(c/s),\log(s/(1-s))$ with Jacobian $(c\,s\,(1-s))^{-1}$. Then, denoting by $\mathsf{P}_{\!\mu,\Sigma}$ the probability measure related to the p.d.f. $h_{\mu,\Sigma}$, the log-likelihood reads
\begin{equation*}
\log\left(L\left((h_{ij \cdot});\mu,\Sigma\right)\right)=\sum_{i=1}^{N_c}\sum_{j=1}^{N_s}h_{ij \cdot}\log\bigg(\mathsf{P}_{\!\mu,\Sigma}\big((c_{i-1},c_i]\times(s_{j-1},s_j]\big)\bigg),
\end{equation*}
where 
\[
h_{ij \cdot}=\sum_{k=1}^{N_{\vartheta}}h_{ijk} 
\]
is the marginal relative frequency in the class $(c_{i-1},c_i]\times(s_{j-1},s_j]$, $i=1,\ldots,N_s$, $j=1,\ldots,N_c$.
%
\section{Quasi-likelihood estimation}\label{sec:QL}
The idea of the quasi-likelihood estimation approach which we aim to apply is to estimate the unknown parameter $\theta$ (taking values in an open subset $\Theta$ of the $p$-dimensional Euclidean space $\mathbb{R}^p$) by finding a root $\hat{\theta}_{QL}$ of the quasi-score estimating function
\begin{equation}\label{eqn:QuasiScore}
Q(\theta,y)=\left(\frac{\partial\mathsf{E}_{\theta}[T(X)]}{\partial\theta}\right)^{\!\!\top}\mathsf{Var}_{\theta}[T(X)]^{-1}(y-\mathsf{E}_{\theta}[T(X)]),
\end{equation}
where $X$ is a random variable on the sample space $\mathbb{X}$ with distribution $\mathsf{P}_{\!\theta}$, $T\colon\mathbb{X}\to\mathbb{R}^q$ is
a transformation of $X$ to a vector of summary statistics, $y=T(x)$ is the respective (column) vector of summary statistics for the observed data $x$, $(\cdot)^\top$ denotes transpose, and, respectively, $\mathsf{E}_{\theta}$ and $\mathsf{Var}_{\theta}$ denote expectation and variance w.r.t. $\mathsf{P}_{\!\theta}$. For a fixed choice $T$ of summary statistics, the quasi-score estimating function $Q$ in (\ref{eqn:QuasiScore}) is that standardized estimation function
\begin{equation}\label{eqn:Standardized}
\tilde{G}=-\left(\mathsf{E}_{\theta}\left[\frac{\partial G}{\partial\theta}\right]\right)^{\!\!\top}\left(\mathsf{E}_{\theta}\left[GG^{\top}\right]\right)^{-1}G,
\end{equation}
$G(\theta,y)=y-\mathsf{E}_{\theta}(T(X))$, for which the information criterion
\begin{equation}\label{eqn:optimality}
\mathcal{E}(G) = \mathsf{E}_{\theta}[\tilde{G}\tilde{G}^{\top}] = \left(\mathsf{E}_{\theta} \left[\frac{\partial G}{\partial\theta}\right]\right)^{\!\!\top}\left(\mathsf{E}_{\theta}\left[GG^{\top}\right]\right)^{-1}\left(\mathsf{E}_{\theta}\left[ \frac{\partial G}{\partial\theta}\right]\right)
\end{equation}
is maximized in the partial order of non-negative definite matrices \citep{ref:GodambeHeyde1987,ref:Heyde1997} among all \emph{linear} standardized unbiased estimating functions of the form $A(\theta)(y-\mathsf{E}_{\theta}(T(X)))$, $A(\theta)$ being any $(p\times q)$-dimensional matrix such that the matrix $A(\theta)\mathsf{Var}_{\theta}[T(X)]A(\theta)^{\top}$ is non-singular.

The information criterion in (\ref{eqn:optimality}) is a generalization of the well-known Fisher information since it coincides with the Fisher information in case a likelihood is available and $G$ equals the usual score function. Then,
in analogy to maximum likelihood estimation, the inverse of $\mathcal{E}(G)$ has
a direct interpretation as the asymptotic variance of the estimator
$\hat{\theta}_{QL}$. For the quasi-likelihood estimation based on (\ref{eqn:QuasiScore}) and the particular vector $T$ of summary statistics the information criterion reads
\begin{equation}\label{eqn:QI}
I_T(\theta)=\mathsf{Var}_{\theta}[Q(\theta,T(X))]=\left(\frac{\partial\mathsf{E}_{\theta}[T(X)]}{\partial\theta}\right)^{\!\!\top}\mathsf{Var}_{\theta}[T(X)]^{-1}\left(\frac{\partial\mathsf{E}_{\theta}[T(X)]}{\partial\theta}\right)
\end{equation}
and is called \emph{quasi-information matrix} in what follows.

Similar to finding a root of the score in maximum likelihood estimation by Fisher scoring, the quasi-score equation (\ref{eqn:QuasiScore}) can be solved with the \emph{Fisher quasi-scoring iteration}
\begin{equation}\label{eqn:QSiteration}
\theta^{(k+1)}=\theta^{(k)}+t^{(k)}\delta^{(k)},\quad\delta^{(k)}=I_T(\theta^{(k)})^{-1}Q(\theta^{(k)},y),
\end{equation}
where $t^{(k)}$ is some step length parameter \citep{ref:Osborne1992,ref:BaaskeEtAl2014}, in case the quasi-score $Q$ and the quasi-information matrix $I_T$ are available as a closed form expression which can be evaluated at least numerically. In particular, this would include to know expectations and variances of the employed summary statistics $T$ w.r.t. to $\mathrm{P}_{\!\theta}$ as a function of $\theta$. However, in many cases including the setting under investigation these expectations and variances are available only as Monte Carlo estimates based on simulated realizations of the random variable $X$ under $\mathrm{P}_{\!\theta}$. Then, still, (\ref{eqn:QSiteration}) might be applied in case simulations are fast and thus the Monte Carlo error can be made small by a sufficiently large number of used model realizations. For more involved simulations, however, a Monte Carlo error cannot be avoided, making a direct application of (\ref{eqn:QSiteration}) unreasonable. In \citet{ref:BaaskeEtAl2014} an idea is presented how the quasi-likelihood estimation approach can be applied even in that case. A respective routine has been developed within the contributed \textsf{R} package \textsf{qle} \citep{ref:Baaske2018}.
%
\section{Simulation study}\label{sec:sim}
The generation of the data for the simulation study includes the following steps. First a realization of a stationary Poisson spheroid process with some fixed intensity $\lambda_V$ (below with the concrete value $50$) and with distribution $\mathrm{P}_{\!\theta}$, $\theta=(\mu_1,\mu_2,\sigma_1,\sigma_2,\varrho,\beta)$, related to the typical (prolate) spheroid (see Section \ref{sec:model}) is generated. Since in this model we allow for spheroids which are not almost surely bounded, it is not sufficient to simulate the Poisson spheroid process in an enlarged window in order to avoid edge effects, see \citet[Sect. 3.1.4]{ref:CSKM2013}. Rather we need an exact simulation method. From the two available general approaches given in \citet[Sect. 13.2.1]{ref:Lantuejoul2002} and \citet{ref:Lantuejoul2013} the one in the latter reference is appropriate but has to be tailored to our specific model. We postpone the details to the Appendix. Then, all spheroids hitting the planar observation window (taken below as a square of side length $10$) inside the section plane $\mathcal{H}$ are identified and the corresponding intersection ellipses are determined. In the sense of an edge correction only section ellipses are considered which have their centres inside the planar observation window. The simulated data then consists of a random number $n$ of triples $(C_l,S_l,\alpha_l)$, $l\in J=\{1,\ldots,n\}$.

For the \emph{unfolding} procedure we consider several grades of fineness for binning the ranges of the 2D data. Since binning the data is a particular case of a kernel density estimation the known problem of small biases and large variances for small bin widths as well as large biases and small variances for large bin widths applies, and the optimal bin width minimizing the mean squared error is unknown. The choices for the numbers $(N_C,N_S,N_{\alpha})$ of bins, ordered from coarse to fine binning, are
\begin{equation}\label{eqn:binning}
(6,5,6),\quad(8,5,8),\quad(12,7,10),\quad(15,10,12),\quad(18,12,15).
\end{equation}
After unfolding the parameters are estimated by maximum likelihood according to the likelihoods given in Section \ref{sec:ML}. These in total five methods are denoted as `UMLE$_1$', `UMLE$_2$', `UMLE$_3$', `UMLE$_4$' and `UMLE$_5$' in the order of the grade of fineness for the binning as in (\ref{eqn:binning}).

Furthermore, writing `med' for the \emph{median}, `mad' for the \emph{median absolute deviation} and `cor' for \emph{Pearson's correlation coefficient}, we employ for the \emph{quasi-likelihood estimation}, denoted by `QLE', the $q=12$ statistics
\begin{align*}
&\mathrm{med}\big(\{\log(C_l)\}_{l\in J}\big),\;\mathrm{med}\big(\{\log(A_l)\}_{l\in J}\big),\;\mathrm{med}\big(\{Y_l\}_{l\in J}\big),\;\mathrm{med}\big(\{S_l\}_{l\in J}\big),\;\mathrm{med}\big(\{\alpha_l\}_{l\in J}\big)\\
&\mathrm{mad}\big(\{\log(C_l)\}_{l\in J}\big),\;\mathrm{mad}\big(\{\log(A_l)\}_{l\in J}\big),\;\mathrm{mad}\big(\{Y_l\}_{l\in J}\big),\;\mathrm{mad}\big(\{S_l\}_{l\in J}\big),\;\mathrm{mad}\big(\{\alpha_l\}_{l\in J}\big)\\
&\mathrm{cor}\big(\{\log(C_l)\}_{l\in J},\{\log(A_l)\}_{l\in J}\big),\;\mathrm{cor}\big(\{\log(A_l)\}_{l\in J},\{Y_l\}_{l\in J}\big),
\end{align*}
where $A_l=C_l/S_l$ and $Y_l=\log(S_l/(1-S_l))$, $l\in J$.

In order not only to have a comparison between the six methods applied to the 2D data but to assess somehow the precision of the respective estimates we have considered estimating the parameters from 3D data, that is, for the case that the spheroids were observable directly and not through a planar section. Besides \emph{maximum likelihood estimation} (denoted by `MLE3D') applied to the 3D data $(c_l,s_l,\vartheta_l)$ we also employ \emph{maximum likelihood estimation} as in Section \ref{sec:ML} \emph{after binning} the 3D data according to the same classification (\ref{eqn:binning}) as used for the 2D data, denoting these methods by `BINMLE$_1$', `BINMLE$_2$', `BINMLE$_3$', `BINMLE$_4$' and `BINMLE$_5$'. Since the latter six methods are based on a different kind of data as the first six methods, each a sub-sample of size $n$ (which is the respective 2D sample size) was taken.  

With each 100 repetitions we have considered five different combinations of parameters. In the first setting the parameters were chosen as
\[
\mu_1=-2.15,\;\mu_2=0.55,\;\sigma_1=0.35,\;\sigma_2=0.3,\;\varrho=0,\;\beta=1,
\]
that is, with independent size and shape and isotropic orientation distribution. In the second and third setting all parameters but $\beta$ are kept the same, and the choices $\beta=10$ and $\beta=0.5$ lead to a relatively strong alignment of the spheroids parallel respectively to a plane, or, to a line, see the reasoning in Section \ref{sec:model} around Eq. (\ref{eqn:pbeta}). In the forth and fifth setting all parameters but $\varrho$ are the same as in the first setting, and the choices $\varrho=0.25$ and $\varrho=0.75$, respectively, imply two different levels of dependence between size and shape. 
\begin{figure}
\centering
\includegraphics[width=0.32\textwidth]{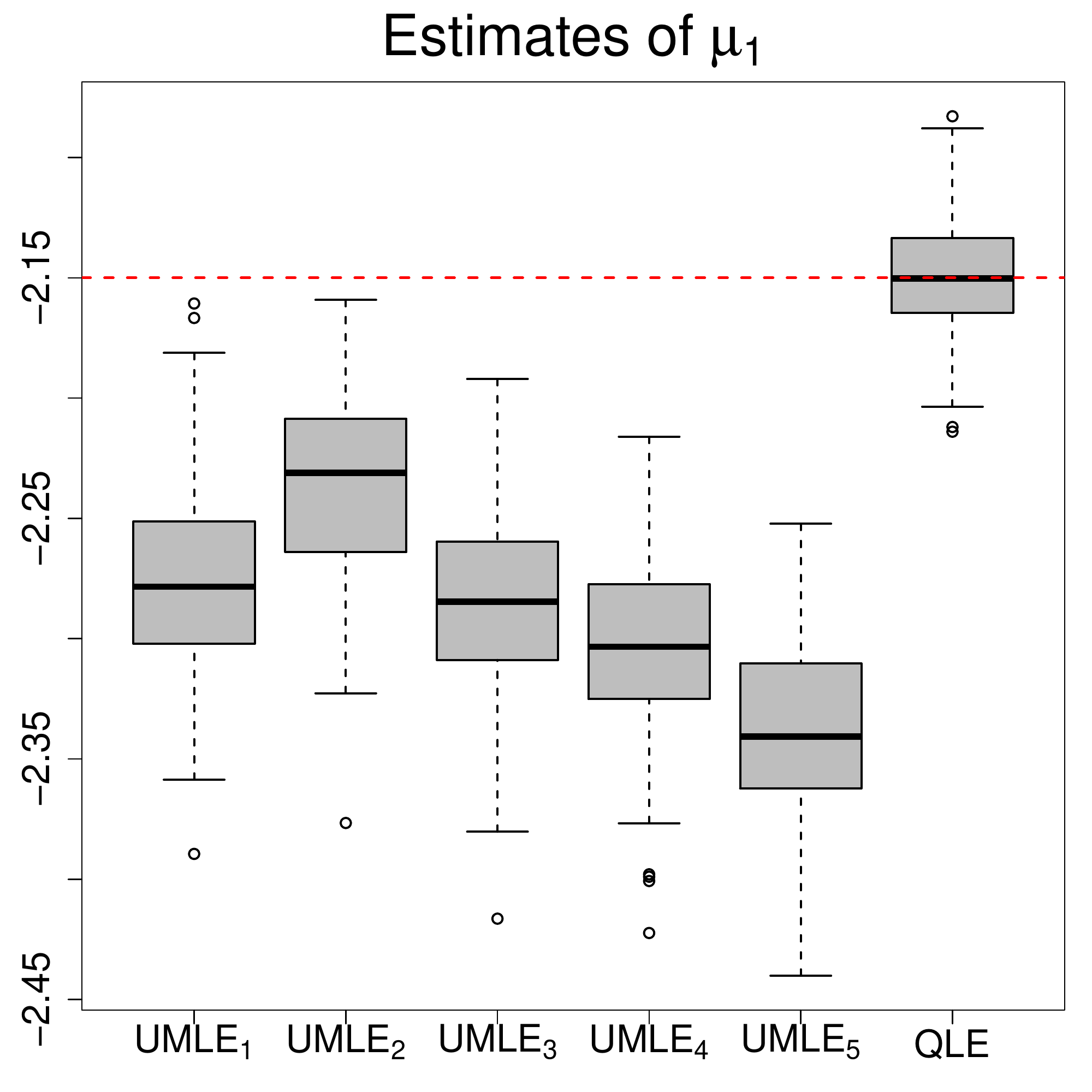}
\includegraphics[width=0.32\textwidth]{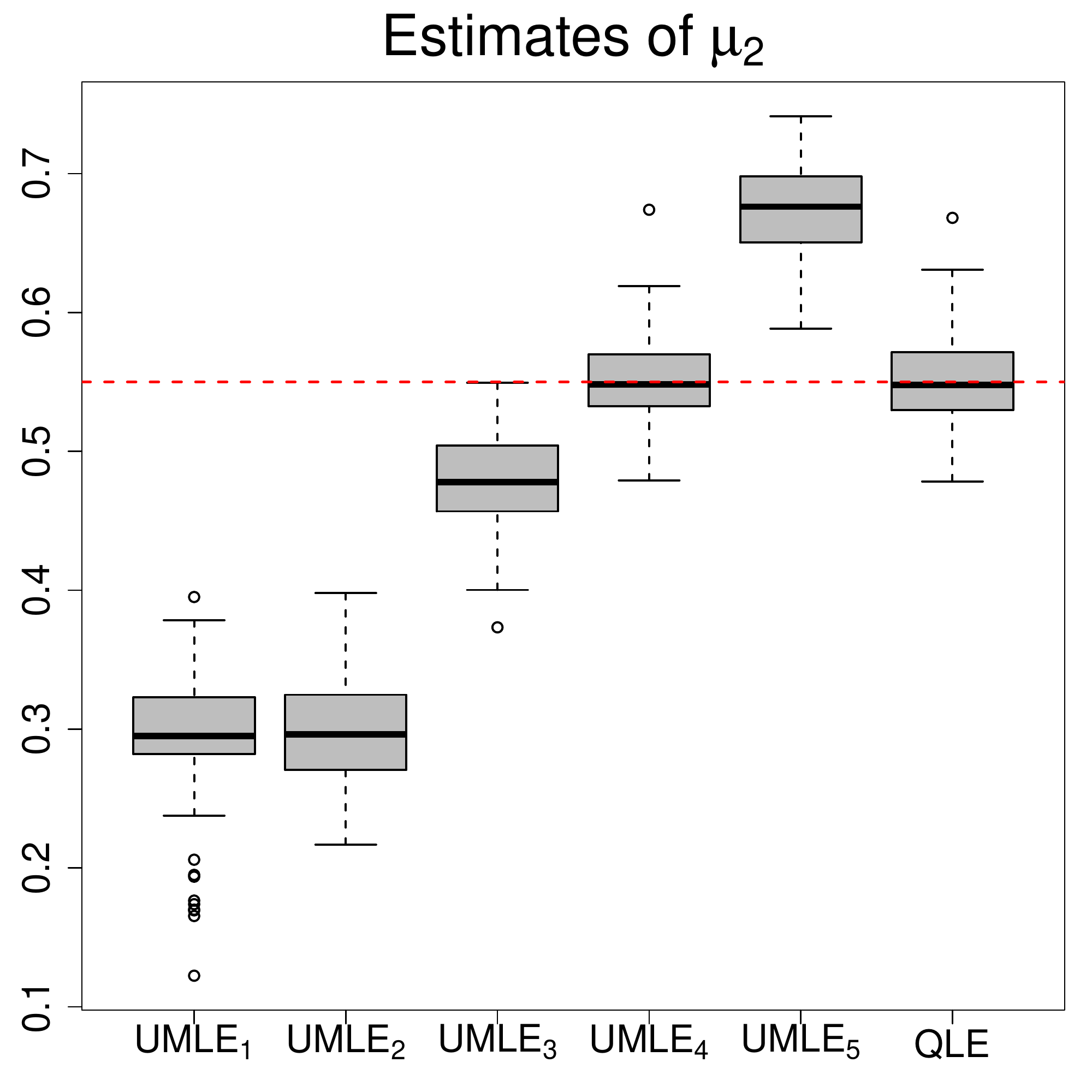}
\includegraphics[width=0.32\textwidth]{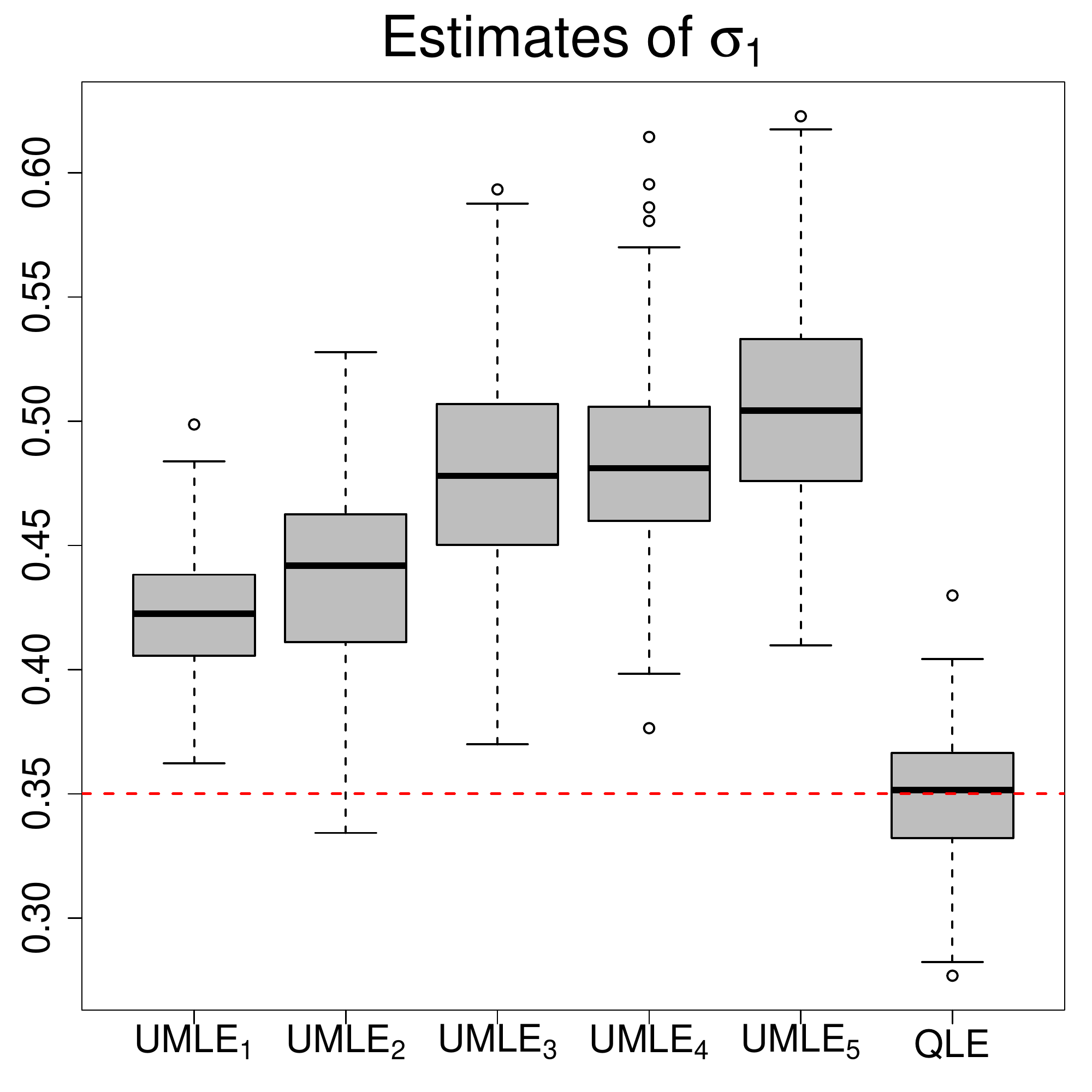}\\
\includegraphics[width=0.32\textwidth]{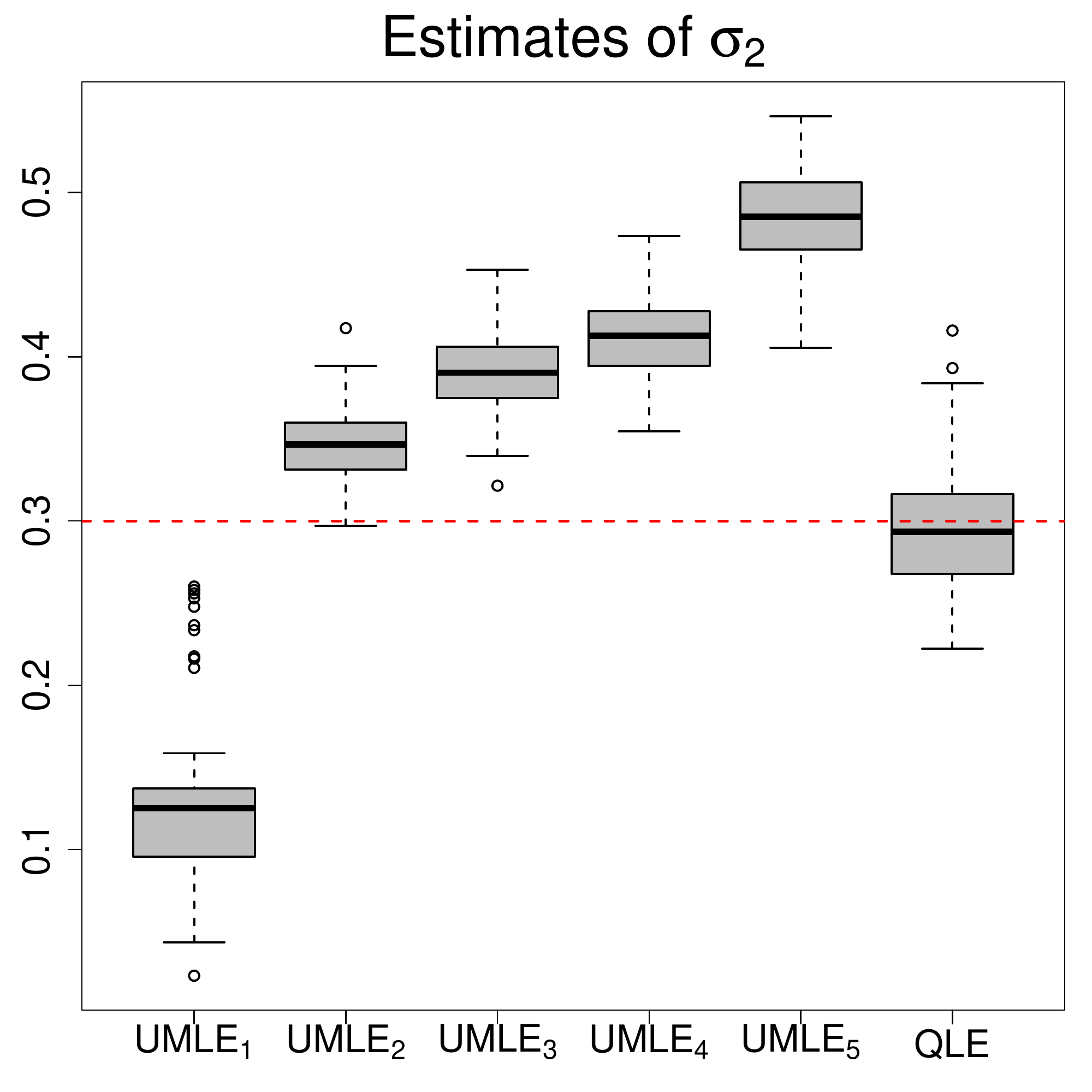}
\includegraphics[width=0.32\textwidth]{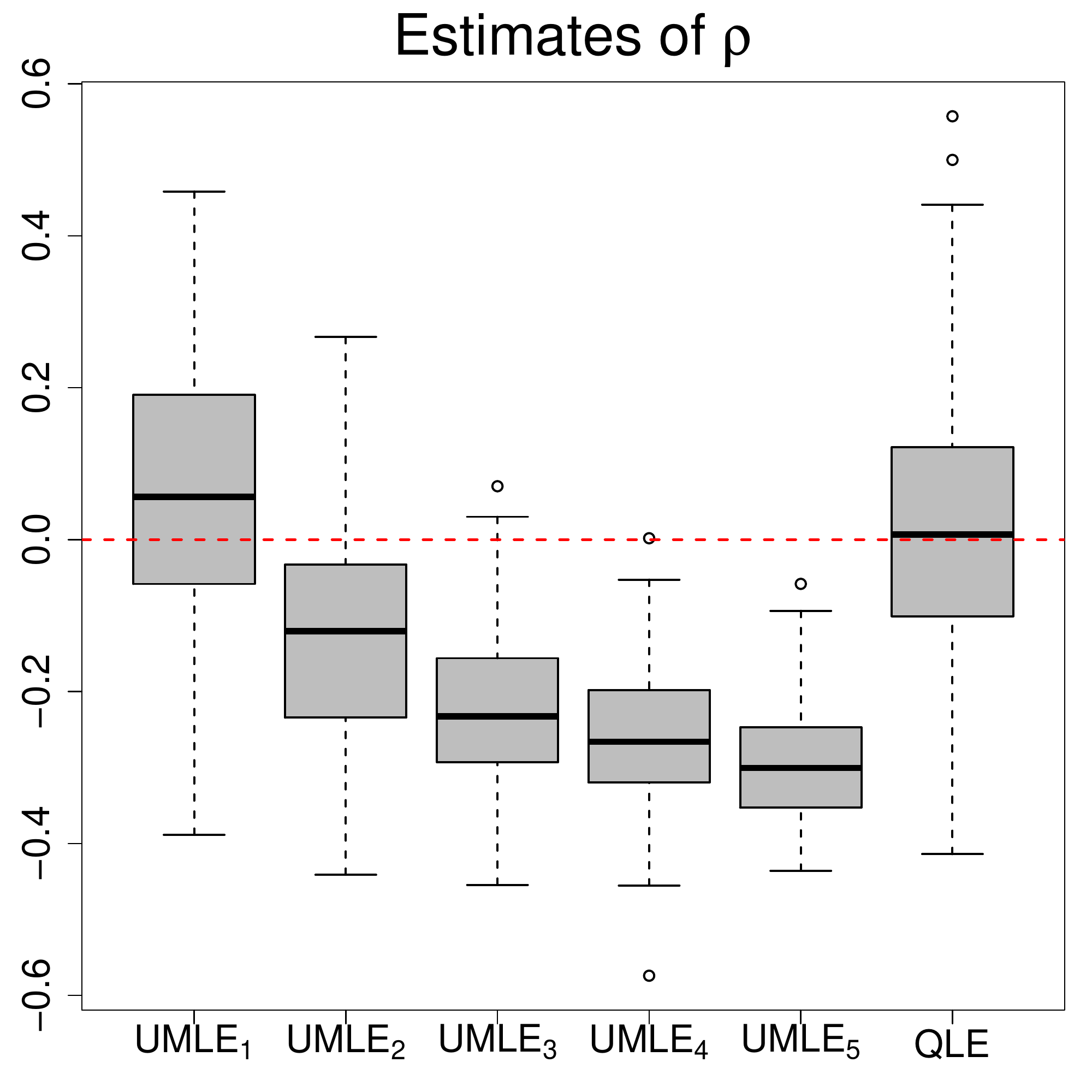}
\includegraphics[width=0.32\textwidth]{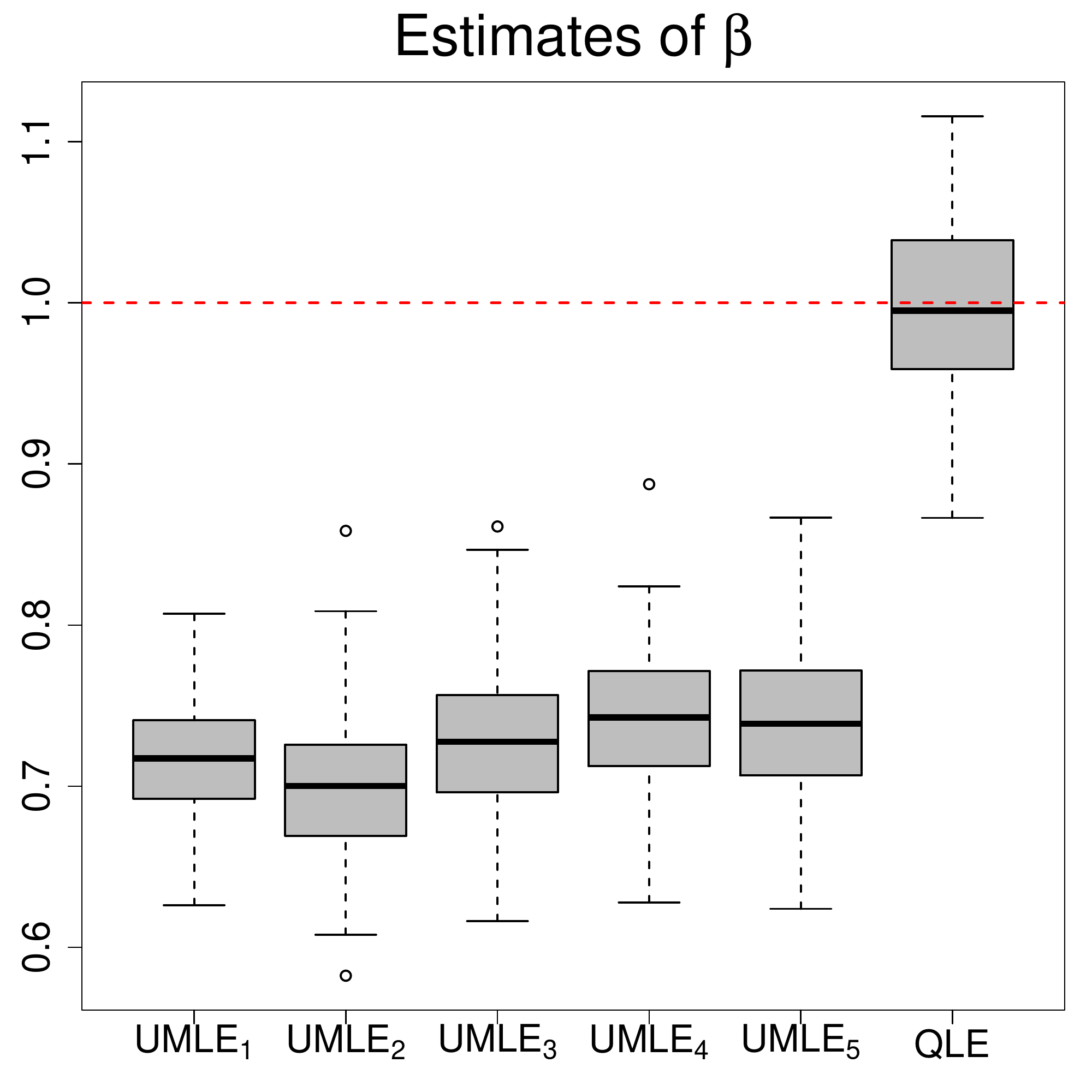}
\caption{Boxplots of the parameter estimates for original parameters $(\mu_1,\mu_2,\sigma_1,\sigma_2,\varrho,\beta)=(-2.15,0.55,0.35,0.3,0.0,1.0)$ (indicated by red dashed lines) from 2D sections with unfolding based on five different bin widths followed each by maximum likelihood estimation (UMLE$_1$,\ldots,UMLE$_5$) as well as with quasi-likelihood estimation (QLE).}
\label{fig:boxplot1}
\end{figure}
\begin{table}
\centering
\caption{Root mean squared error (rmse) of the parameter estimates for original parameters $(\mu_1,\mu_2,\sigma_1,\sigma_2,\varrho,\beta)=(-2.15,0.55,0.35,0.3,0.0,1.0)$ (top) from 2D sections with unfolding based on five different bin widths followed each by maximum likelihood estimation (UMLE$_1$,\ldots,UMLE$_5$) as well as with quasi-likelihood estimation (QLE), and (bottom) from 3D data with maximum likelihood (MLE3D) and with maximum likelihood after classifying the 3D data into the same five binnings as used for unfolding (BINMLE$_1$,\ldots,BINMLE$_5$).}
\label{tab:rmse1}
\begin{tabular}{l|rrrrrr}
Method&rmse$(\hat{\mu}_1)$&rmse$(\hat{\mu}_2)$&rmse$(\hat{\sigma}_1)$&rmse$(\hat{\sigma}_2)$&rmse$(\hat{\varrho})$&rmse$(\hat{\beta}$)\\
\hline
UMLE$_1$&0.133&0.259&0.077&0.184&0.206&0.285\\
ULME$_2$&0.096&0.254&0.095&0.052&0.193&0.303\\
ULME$_3$&0.142&0.080&0.137&0.094&0.247&0.276\\
ULME$_4$&0.159&0.032&0.142&0.114&0.274&0.262\\
ULME$_5$&0.195&0.128&0.164&0.187&0.306&0.264\\
\hline
QLE&0.025&0.033&0.027&0.040&0.167&0.053
\end{tabular}

\bigskip

\begin{tabular}{l|rrrrrr}
Method&rmse$(\hat{\mu}_1)$&rmse$(\hat{\mu}_2)$&rmse$(\hat{\sigma}_1)$&rmse$(\hat{\sigma}_2)$&rmse$(\hat{\varrho})$&rmse$(\hat{\beta}$)\\
\hline
BINMLE$_1$&0.013&0.041&0.012&0.087&0.088&0.036\\
BINMLE$_2$&0.011&0.011&0.011&0.008&0.032&0.039\\
BINMLE$_3$&0.013&0.011&0.010&0.008&0.038&0.032\\
BINMLE$_4$&0.013&0.011&0.008&0.008&0.037&0.034\\
BINMLE$_5$&0.013&0.010&0.008&0.007&0.035&0.035\\
\hline
MLE3D&0.012&0.011&0.008&0.007&0.035&0.039
\end{tabular}
\end{table}
\begin{table}
\centering
\caption{Root mean squared error (rmse) of the parameter estimates for original parameters $(\mu_1,\mu_2,\sigma_1,\sigma_2,\varrho,\beta)=(-2.15,0.55,0.35,0.3,0.0,10.0)$ (top) from 2D sections with unfolding based on five different bin widths followed each by maximum likelihood estimation (UMLE$_1$,\ldots,UMLE$_5$) as well as with quasi-likelihood estimation (QLE), and (bottom) from 3D data with maximum likelihood (MLE3D) and with maximum likelihood after classifying the 3D data into the same five binnings as used for unfolding (BINMLE$_1$,\ldots,BINMLE$_5$).}
\label{tab:rmse2}
\begin{tabular}{l|rrrrrr}
Method&rmse$(\hat{\mu}_1)$&rmse$(\hat{\mu}_2)$&rmse$(\hat{\sigma}_1)$&rmse$(\hat{\sigma}_2)$&rmse$(\hat{\varrho})$&rmse$(\hat{\beta}$)\\
\hline
UMLE$_1$&0.119&0.269&0.093&0.203&0.256&6.382\\
ULME$_2$&0.069&0.298&0.097&0.038&0.182&6.399\\
ULME$_3$&0.130&0.039&0.127&0.225&0.302&5.972\\
ULME$_4$&0.145&0.073&0.139&0.255&0.322&5.409\\
ULME$_5$&0.183&0.233&0.155&0.363&0.353&5.125\\
\hline
QLE&0.039&0.066&0.031&0.050&0.284&0.459
\end{tabular}

\bigskip

\begin{tabular}{l|rrrrrr}
Method&rmse$(\hat{\mu}_1)$&rmse$(\hat{\mu}_2)$&rmse$(\hat{\sigma}_1)$&rmse$(\hat{\sigma}_2)$&rmse$(\hat{\varrho})$&rmse$(\hat{\beta}$)\\
\hline
BINMLE$_1$&0.012&0.032&0.014&0.062&0.065&0.792\\
BINMLE$_2$&0.013&0.011&0.010&0.008&0.037&0.707\\
BINMLE$_3$&0.011&0.009&0.009&0.007&0.035&0.641\\
BINMLE$_4$&0.010&0.011&0.008&0.008&0.040&0.488\\
BINMLE$_5$&0.013&0.009&0.008&0.007&0.032&0.434\\
\hline
MLE3D&0.010&0.009&0.007&0.006&0.033&0.370
\end{tabular}
\end{table}
\begin{table}
\centering
\caption{Root mean squared error (rmse) of the parameter estimates for original parameters $(\mu_1,\mu_2,\sigma_1,\sigma_2,\varrho,\beta)=(-2.15,0.55,0.35,0.3,0.0,0.5)$ (top) from 2D sections with unfolding based on five different bin widths followed each by maximum likelihood estimation (UMLE$_1$,\ldots,UMLE$_5$) as well as with quasi-likelihood estimation (QLE), and (bottom) from 3D data with maximum likelihood (MLE3D) and with maximum likelihood after classifying the 3D data into the same five binnings as used for unfolding (BINMLE$_1$,\ldots,BINMLE$_5$).}
\label{tab:rmse3}
\begin{tabular}{l|rrrrrr}
Method&rmse$(\hat{\mu}_1)$&rmse$(\hat{\mu}_2)$&rmse$(\hat{\sigma}_1)$&rmse$(\hat{\sigma}_2)$&rmse$(\hat{\varrho})$&rmse$(\hat{\beta}$)\\
\hline
UMLE$_1$&0.114&0.287&0.08&0.186&0.197&0.106\\
ULME$_2$&0.078&0.306&0.097&0.029&0.182&0.131\\
ULME$_3$&0.114&0.150&0.137&0.074&0.257&0.124\\
ULME$_4$&0.137&0.077&0.148&0.090&0.272&0.121\\
ULME$_5$&0.168&0.046&0.162&0.147&0.293&0.123\\
\hline
QLE&0.027&0.032&0.026&0.034&0.163&0.027
\end{tabular}

\bigskip

\begin{tabular}{l|rrrrrr}
Method&rmse$(\hat{\mu}_1)$&rmse$(\hat{\mu}_2)$&rmse$(\hat{\sigma}_1)$&rmse$(\hat{\sigma}_2)$&rmse$(\hat{\varrho})$&rmse$(\hat{\beta}$)\\
\hline
BINMLE$_1$&0.015&0.048&0.012&0.101&0.097&0.018\\
BINMLE$_2$&0.012&0.013&0.011&0.010&0.039&0.019\\
BINMLE$_3$&0.011&0.011&0.010&0.009&0.037&0.019\\
BINMLE$_4$&0.012&0.011&0.011&0.008&0.038&0.017\\
BINMLE$_5$&0.011&0.010&0.010&0.007&0.035&0.017\\
\hline
MLE3D&0.011&0.010&0.009&0.007&0.035&0.017
\end{tabular}
\end{table}
\begin{table}
\centering
\caption{Root mean squared error (rmse) of the parameter estimates for original parameters $(\mu_1,\mu_2,\sigma_1,\sigma_2,\varrho,\beta)=(-2.15,0.55,0.35,0.3,0.25,1.0)$ from 2D sections with unfolding based on five different bin widths followed each by maximum likelihood estimation (UMLE$_1$,\ldots,UMLE$_5$) as well as with quasi-likelihood estimation (QLE).}
\label{tab:rmse4}
\begin{tabular}{l|rrrrrr}
Method&rmse$(\hat{\mu}_1)$&rmse$(\hat{\mu}_2)$&rmse$(\hat{\sigma}_1)$&rmse$(\hat{\sigma}_2)$&rmse$(\hat{\varrho})$&rmse$(\hat{\beta}$)\\
\hline
UMLE$_1$&0.160&0.239&0.083&0.202&0.224&0.285\\
ULME$_2$&0.121&0.231&0.098&0.051&0.262&0.302\\
ULME$_3$&0.161&0.069&0.148&0.094&0.406&0.277\\
ULME$_4$&0.179&0.040&0.155&0.108&0.431&0.268\\
ULME$_5$&0.211&0.142&0.175&0.185&0.485&0.264\\
\hline
QLE&0.031&0.047&0.032&0.046&0.258&0.050
\end{tabular}
\end{table}
\begin{table}
\centering
\caption{Root mean squared error (rmse) of the parameter estimates for original parameters $(\mu_1,\mu_2,\sigma_1,\sigma_2,\varrho,\beta)=(-2.15,0.55,0.35,0.3,0.75,1.0)$ from 2D sections with unfolding based on five different bin widths followed each by maximum likelihood estimation (UMLE$_1$,\ldots,UMLE$_5$) as well as with quasi-likelihood estimation (QLE).}
\label{tab:rmse5}
\begin{tabular}{l|rrrrrr}
Method&rmse$(\hat{\mu}_1)$&rmse$(\hat{\mu}_2)$&rmse$(\hat{\sigma}_1)$&rmse$(\hat{\sigma}_2)$&rmse$(\hat{\varrho})$&rmse$(\hat{\beta}$)\\
\hline
UMLE$_1$&0.217&0.188&0.081&0.242&0.225&0.287\\
ULME$_2$&0.171&0.174&0.104&0.049&0.444&0.291\\
ULME$_3$&0.195&0.041&0.148&0.084&0.685&0.281\\
ULME$_4$&0.214&0.072&0.158&0.106&0.716&0.273\\
ULME$_5$&0.240&0.179&0.178&0.176&0.786&0.272\\
\hline
QLE&0.034&0.050&0.044&0.084&0.266&0.066
\end{tabular}
\end{table}

The results of the simulation study for the first choice $(-2.15,0.55,0.35,0.3,0.0,1.0)$ of parameters, given in Figure \ref{fig:boxplot1} in terms of boxplots and in Table \ref{tab:rmse1} in terms of root mean squared errors (corresponding bootstrap standard errors are given in Table \ref{tab:sderr1}) show that QLE provides clearly smaller mean squared errors than most of those from UMLE$_1$,\ldots,UMLE$_5$, and that the reason for this behaviour is most of all due to the bias resulting from unfolding. The comparison with the maximum likelihood estimates from the 3D data shows that the effect of binning the data into classes is small whatever the grade of fineness is. Likewise the precision of QLE is roughly (only) one order of magnitude worse than that of MLE3D but one order of magnitude better than that of UMLE$_1$,\ldots,UMLE$_5$. Figure \ref{fig:boxplot1} shows that a particular choice of bins leads for a certain parameter to a precision comparable to that of QLE whereas with the same choice of bins estimation of another parameter implies a certain bias.
These findings also hold for other choices of the parameters, see the root mean squared errors in the Tables \ref{tab:rmse2}--\ref{tab:rmse5} together with the respective bootstrap standard errors in Tables \ref{tab:sderr2}--\ref{tab:sderr5} (postponed to the Appendix). In particular, it becomes obvious from Table \ref{tab:rmse2} that the error of the estimates of $\beta=10$ from UMLE$_1$,\ldots,UMLE$_5$ is almost as large as the parameter. However, in either case QLE as well as UMLE$_1$,\ldots,UMLE$_5$ produce large errors when estimating the parameter $\varrho$, which is not the case when using 3D data, see Tables \ref{tab:rmse1} to \ref{tab:rmse3}.
%
\section{Application}\label{sec:appl}
%
We consider an aluminium matrix composite (see Figure \ref{fig:data}) reinforced with alumina particles (Al$_2$O$_3$, ca. 15\% volume fraction), denoted by AA6061-15p, as in \citet{ref:MuellerEtAl2015} where experimental investigations most of all on the very high cycle fatigue behaviour are performed and further details may be found. Due to extrusion moulding into a bar the particles in AA6061-15p were aligned nearly parallel to the extrusion direction. Therefore, the production process justifies the assumption that the orientation distribution of the reinforcements is invariant with respect to rotations about some fixed axis. 

The ceramic inclusions are modelled as prolate spheroids. Although the investigation of a similar sample of ceramic inclusions in \citet{ref:BorbelyEtAl2004} indicates that a modelling with general ellipses could be even better, prolate spheroids seem to be still sufficiently flexible for the particles and make corresponding planar section data accessible for stereological unfolding in the above sense (Section \ref{sec:unfold}). In order to extract the data for the corresponding ellipses from the planar section the 2D image has been segmented and ellipses have been fitted with the help of an image analyser \citep{ref:PauEtAl2010}; the ellipses data is available as the dataset \textsf{data15p} within the \textsf{R} package \textsf{unfoldr} \citep{ref:Baaske2017}. The aim is to estimate the parameters of the supposed spheroid distribution (see Section \ref{sec:model}) from the 2D data with the quasi-likelihood approach. A section plane parallel to the extrusion direction (see Figure \ref{fig:data}, left) was used which corresponds to a vertical uniform random section as needed for unfolding (see Section \ref{sec:unfold}). 

In order to estimate the statistics (see Section \ref{sec:sim}) needed for QLE an edge correction in the sense of minus-sampling \citep[][p. 254]{ref:CSKM2013} has been applied. The results of QLE are given in Table \ref{tab:sqle}, together with the asymptotic standard errors of these estimates as the entries of $\sqrt{\operatorname{diag}(I(\hat{\theta}_{QL})^{-1})}$, cf. Section 4 and \citet{ref:BaaskeEtAl2014}.
\begin{table}
\centering
\caption{Estimates and asymptotic standard errors after applying QLE to the AA6061-15p data.}
\label{tab:sqle}
\begin{tabular}{l|r|r|r|r|r|r}
parameter&$\mu_1$&$\mu_2$&$\sigma_1$&$\sigma_2$&$\varrho$&$\beta$\\
\hline
estimate&$-3.706$&$-0.782$&$0.358$&$0.567$&$-0.218$&$0.194$\\
\hline
standard error&$0.041$&$0.050$&$0.052$&$0.043$&$0.122$&$0.011$
\end{tabular}
\end{table}

In order to assess the goodness-of-fit of the QLE parameter estimates we performed each a Kolmogorov-Smirnov test for the length $A$ of the semi-major axis, the length $C$ of the semi-minor axis, the shape factor $S$ and the direction angle $\alpha$ of the section ellipses. The corresponding $p$-values are
\begin{equation*}
p_A=0.071,\;p_C=0.095,\;p_S=0.070,\;p_{\alpha}=0.114\,,
\end{equation*}
indicating that the fitted model indeed reflects well the marginal distributions of the essential 2D quantities.

\begin{figure}
\centering
\includegraphics[width=0.4\textwidth]{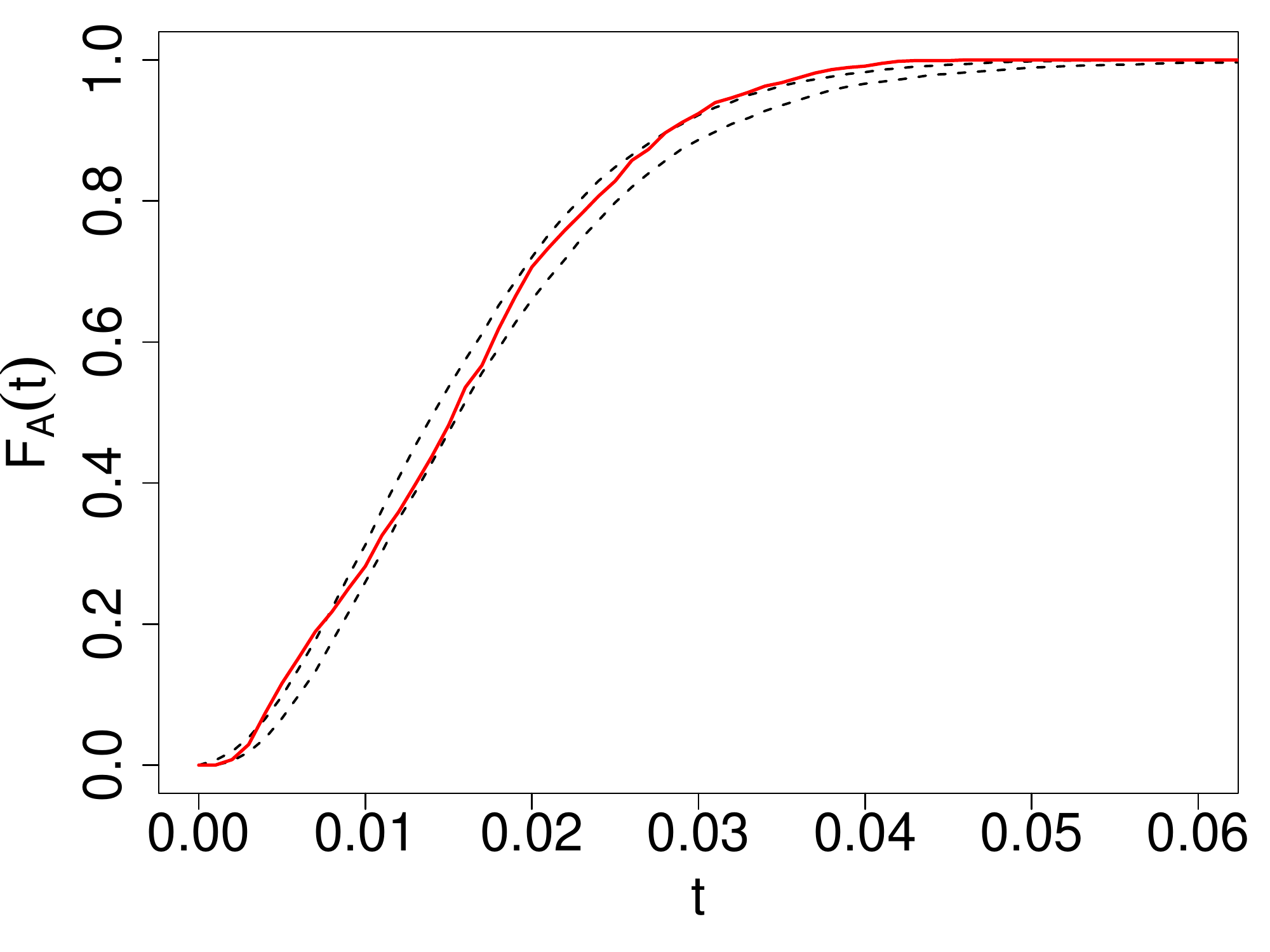}\qquad
\includegraphics[width=0.4\textwidth]{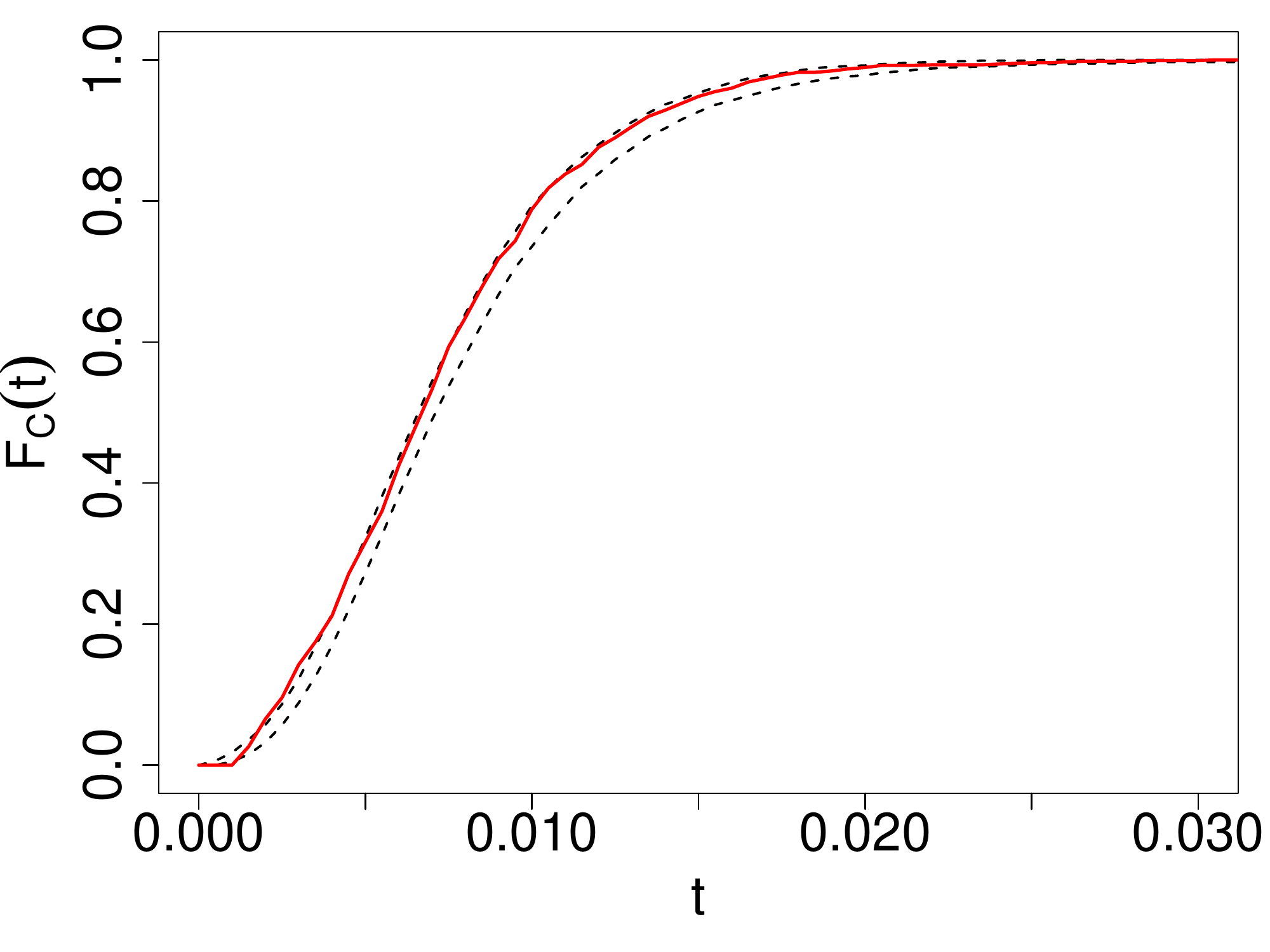}\\[0.2em]
\includegraphics[width=0.4\textwidth]{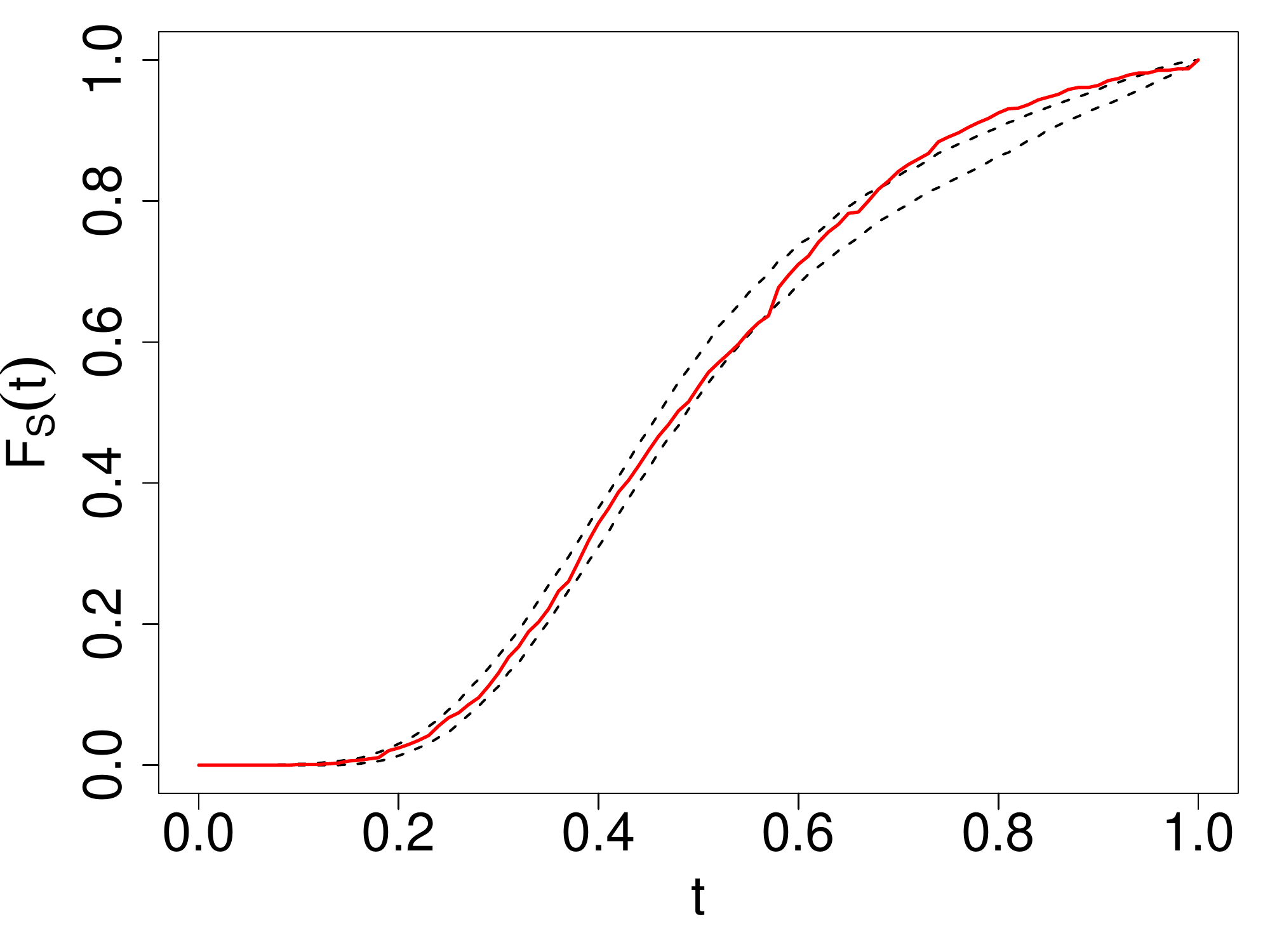}\qquad
\includegraphics[width=0.4\textwidth]{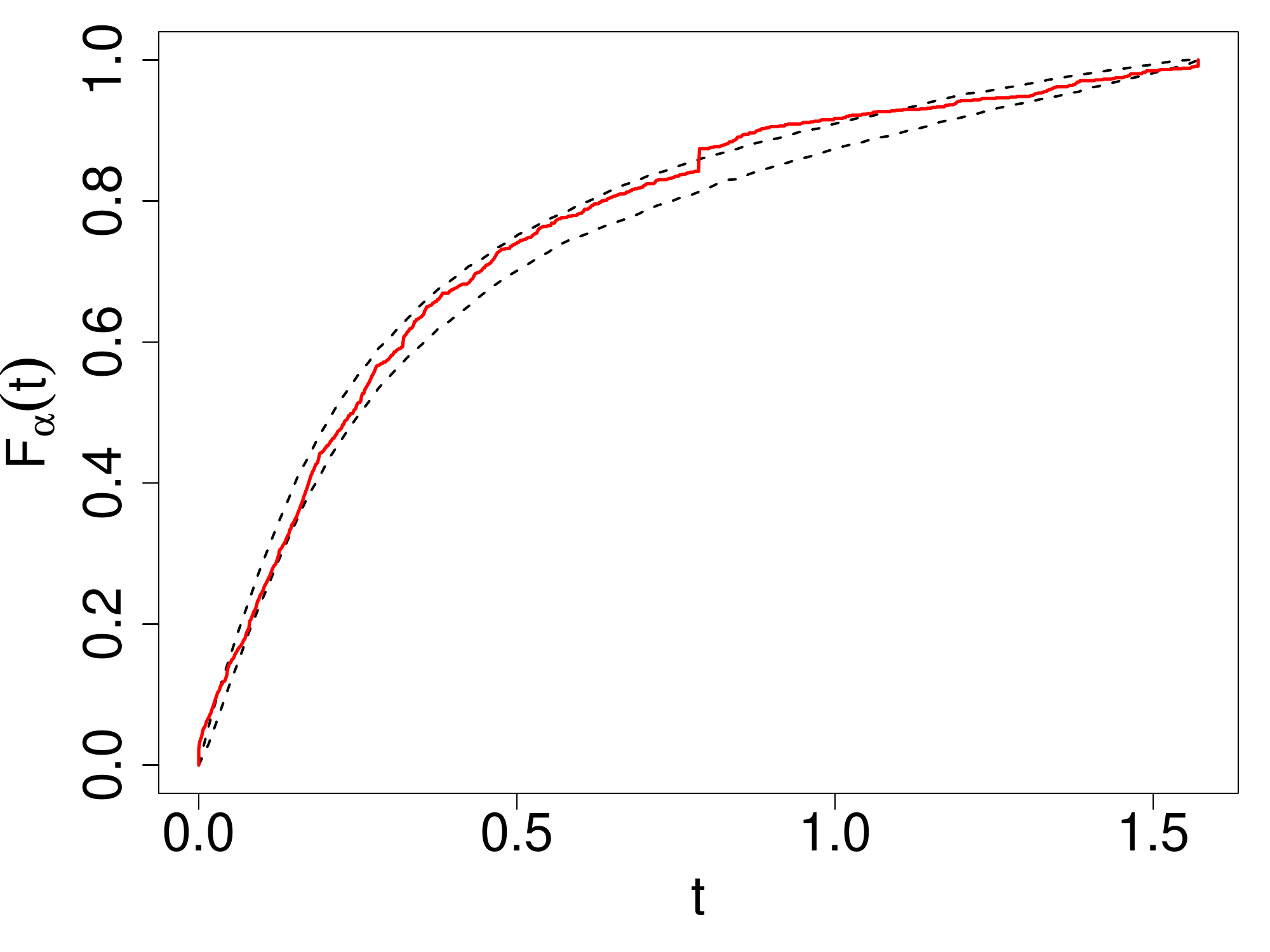}
\caption{Empirical c.d.f. for the AA6061-15p data (red) and pointwise 95\% envelopes of the c.d.f. for the model fitted with quasi-likelihood (black dashed) of length $A$ of the semi-major axis ($F_A$, top left), of length $C$ of the semi-minor axis ($F_C$, top right), of shape factor $S$ ($F_S$, bottom left) and of angle $\alpha$ ($F_{\alpha}$, bottom right).}
\label{fig:sqle:gof15p}
\end{figure}

Furthermore, we generated each 199 samples of a Poisson spheroid system such that the mean number of spheroids hitting the observation window equals the sample size of the data. From the data and from the simulated samples each the empirical c.d.f. of the length $A$ of the semi-major axis, of the length $C$ of the semi-minor axis, of the shape factor $S$ and of the direction angle $\alpha$ of the intersection ellipses were determined. In Figure \ref{fig:sqle:gof15p} the pointwise 95\% envelopes of the empirical c.d.f.s from the simulated samples as well as the corresponding empirical c.d.f. of the data are plotted.
The small jumps in Figure \ref{fig:sqle:gof15p} (bottom right) in the empirical c.d.f. $F_{\alpha}$ w.r.t. to the angle $\alpha$ at 0, $\pi/4$ and $\pi/2$ are clearly due to digitisation. All in all it seems that the QLE fit of the spheroid distribution for AA6061-15p can be considered as satisfying. 

\begin{figure}
\centering
\includegraphics[width=0.4\textwidth]{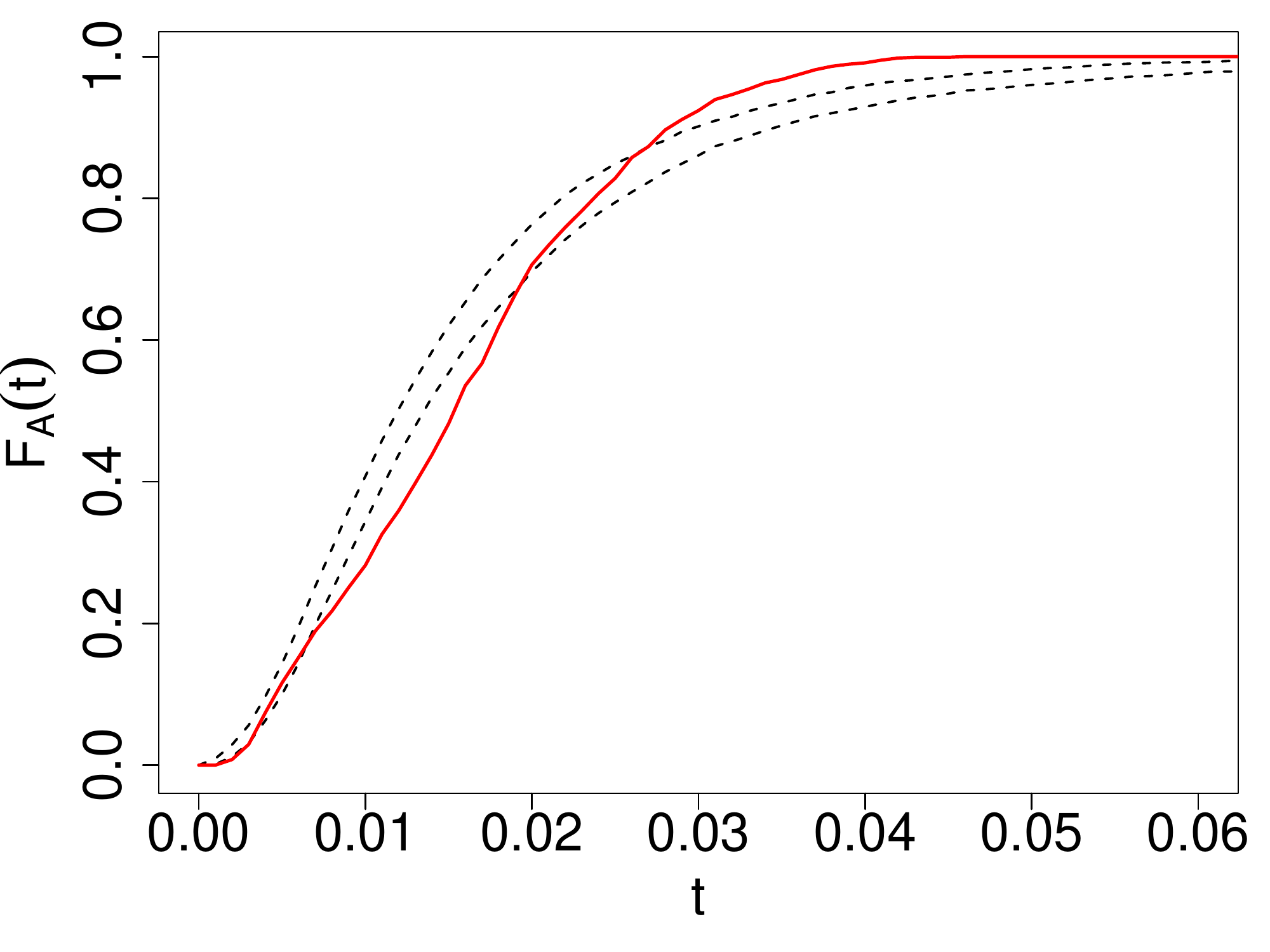}\qquad
\includegraphics[width=0.4\textwidth]{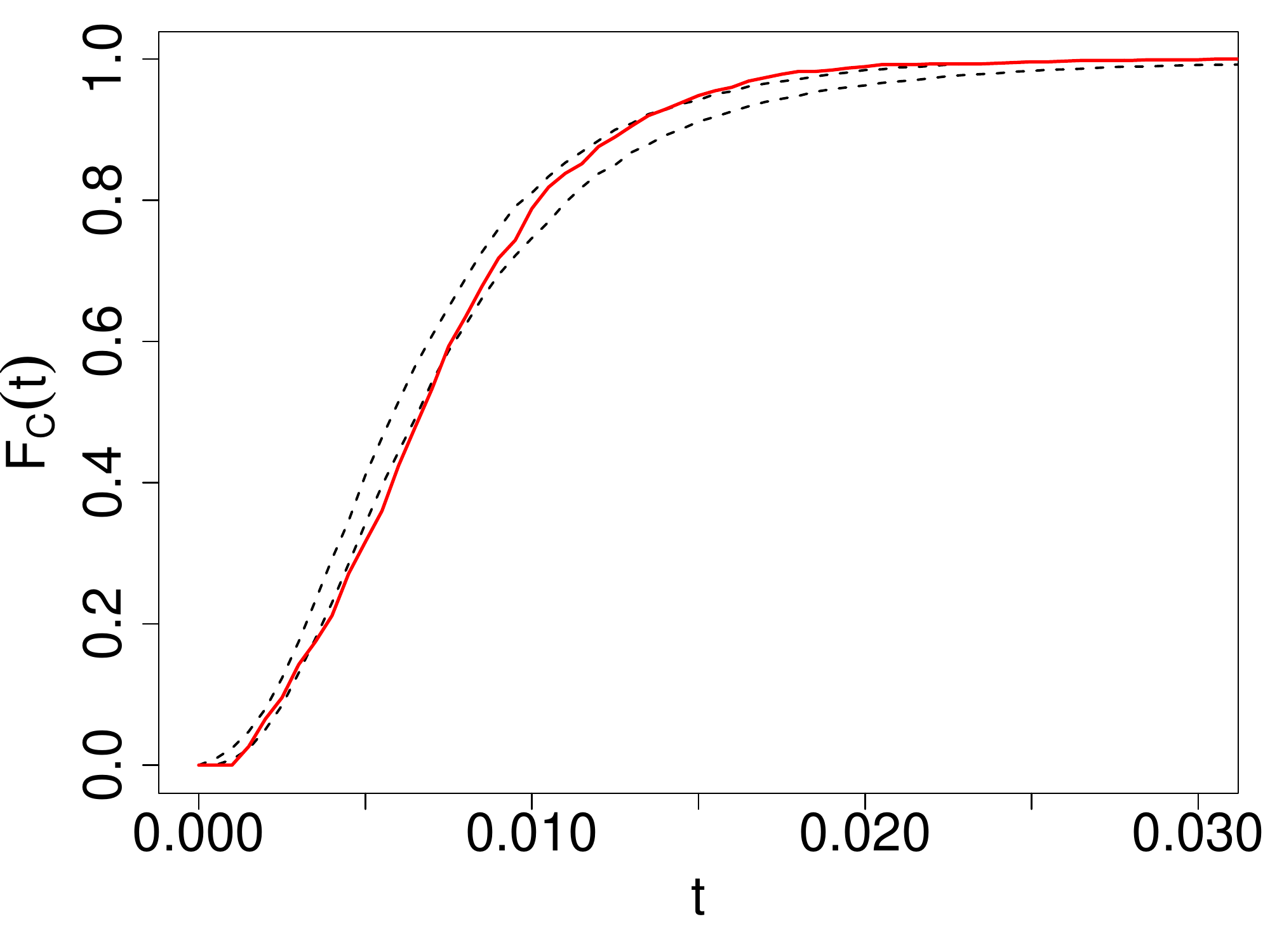}\\[0.2em]
\includegraphics[width=0.4\textwidth]{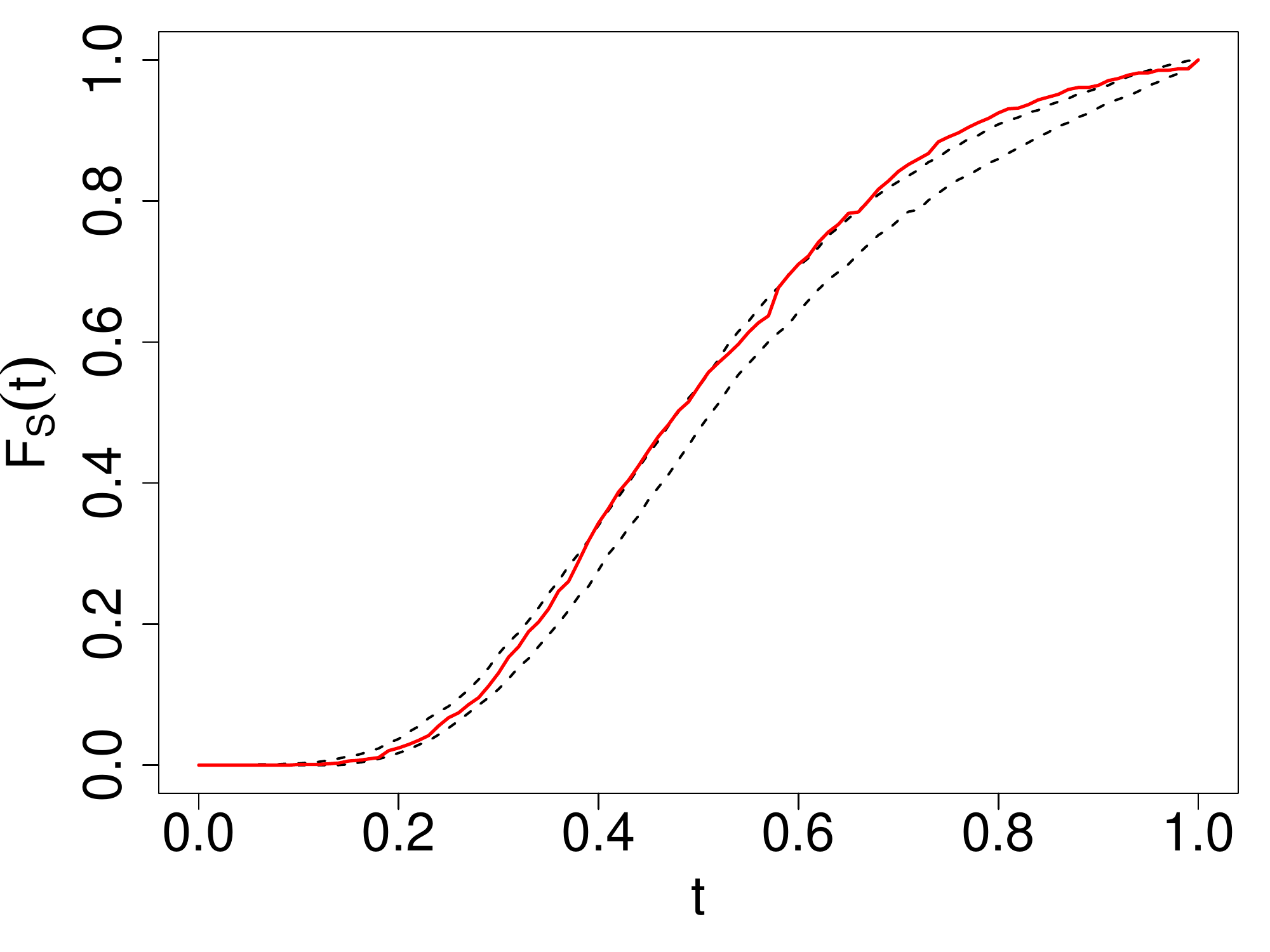}\qquad
\includegraphics[width=0.4\textwidth]{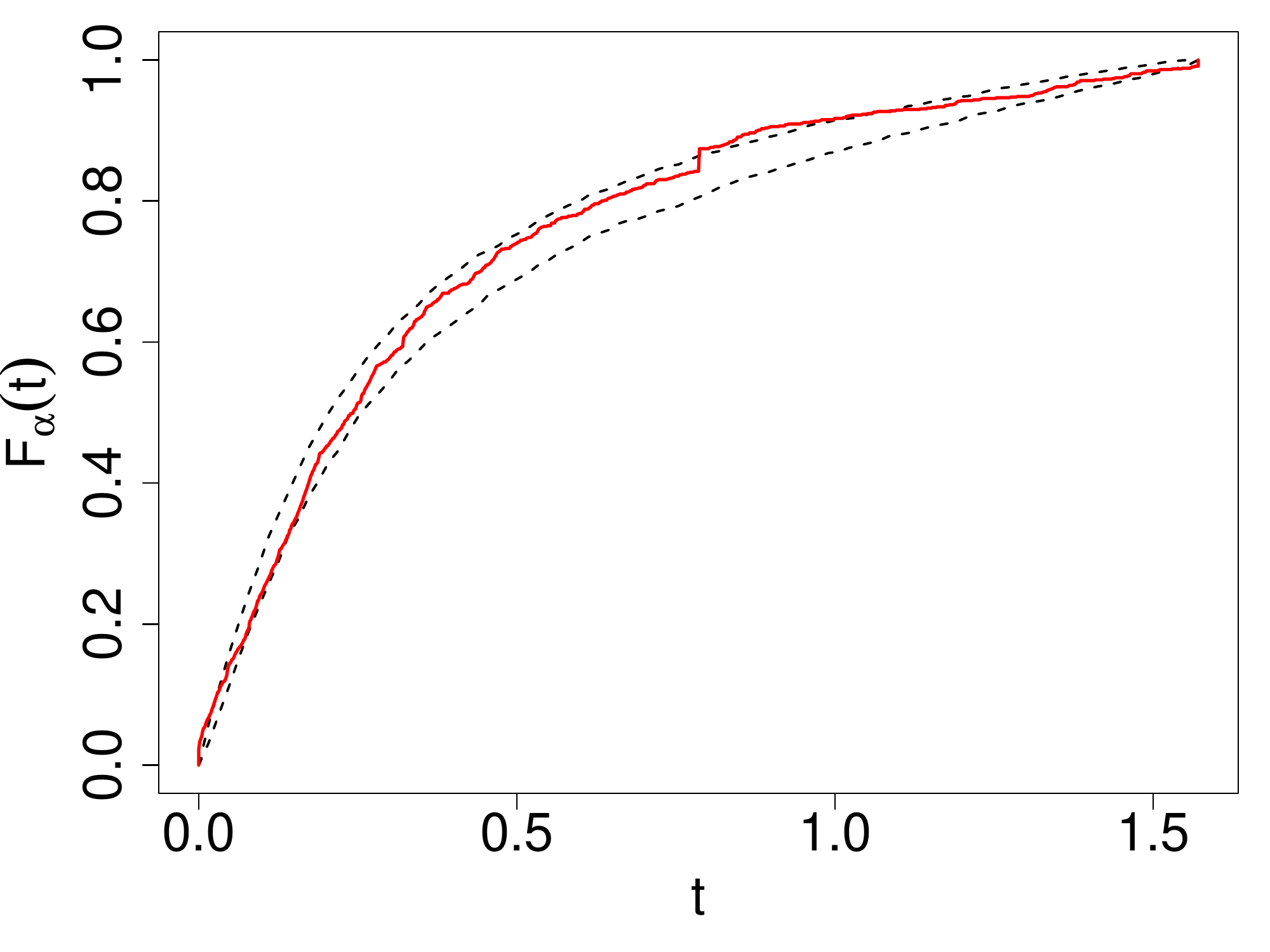}
\caption{Empirical c.d.f. for the AA6061-15p data (red) and pointwise 95\% envelopes of the c.d.f. for the model fitted with maximum likelihood after unfolding (black dashed) of length $A$ of the semi-major axis ($F_A$, top left), of length $C$ of the semi-minor axis ($F_C$, top right), of shape factor $S$ ($F_S$, bottom left) and of angle $\alpha$ ($F_{\alpha}$, bottom right).}
\label{fig:umle:gof15p}
\end{figure}

Although the simulation study in Section \ref{sec:sim} indicates that QLE typically outperforms ULME we also fitted the model with maximum likelihood after unfolding. Among UMLE$_1$,\ldots,UMLE$_5$ the variant UMLE$_4$ performs best and results in the following parameter estimates:
\begin{equation*}
\hat{\mu}_1=-4.030,\;\hat{\mu}_2=-0.584,\;\hat{\sigma}_1=0.637,\;\hat{\sigma}_2=0.677,\;\hat{\varrho}=-0.531,\;\hat{\beta}=0.194
\end{equation*} 
Interestingly, the estimated value of $\beta$ is approximately the same with both methods while there is a clear difference for the other parameter estimates. The Kolmogorov-Smirnov test for $A$, $C$, $S$ and $\alpha$ results in the $p$-values
\begin{equation*}
p_A=0.00001,\;p_C=0.00016,\;p_S=0.04534,\;p_{\alpha}=0.20664\,,
\end{equation*}
such that, even after correcting for multiple comparisons, the model obtained with UMLE$_4$ has to be rejected (cf. also Figure \ref{fig:umle:gof15p}).

%
\section{Conclusions}\label{sec:concl}
We have demonstrated the potential of the quasi-likelihood estimation approach for inferring the parameters of a distribution of three-dimensional objects which are observable only through two-dimensional sections, even in the more complicated situation that objective functions are only available as Monte Carlo approximations. Although for certain bin widths of the histograms the precision of the parameter estimates with the maximum likelihood method after the well-known procedure of unfolding is comparable to the precision of the quasi-likelihood estimates the respective bin widths are not known in advance and, more seriously, in general differ for the different parameters. At least in the case of a \emph{parametric} modelling approach the quasi-likelihood estimation approach should thus be considered as a valuable alternative. For a \emph{non-parametric} modelling approach, however, unfolding will keep indispensable. 

The presented study also strongly supports the so-called `two-step method' as recommended in \citet[][p. 433]{ref:CSKM2013} for parametric inference based on sectional data, namely first to get ideas on the type of distributions in, say, 3D from a rough solution of the unfolding problem and then to apply a parametric approach involving only the sectional data. 

Although the employed parametric model for the random prolate spheroids is of course a particular one, the two investigated methods of parameter estimation might be adapted to other models. We would then expect a similar behaviour w.r.t. the precision of the parameter estimates. Nevertheless, the simulation-based quasi-likelihood estimation approach as implemented in \citet{ref:Baaske2018} is basically applicable in very general settings and thus a powerful tool for parameter estimation in spatial statistics.

For the real-world sample in Section \ref{sec:appl} it turns out that the particular parametric spheroid model obtained with the quasi-likelihood approach is able to model the distribution of the ceramic particles in AA6061-15p reasonable well. In particular, the goodness-of-fit test results indicate that the restriction to spheroids, i.\,e. ellipsoids of revolution, as the shape model for the ceramic inclusions is sufficient. Hence, the fitted model can then be further used, most of all for generating virtual particle configurations for assessing the very high cycle fatigue behaviour of metal-matrix composites as in \citet{ref:BaaskeEtAl2018Book}. 
\section*{Acknowledgments}
Support of the research by the German Research 
Foundation (DFG) within the priority programme 
``Life${}^{\infty}$'' (SPP 1466) is gratefully acknowledged.
%
\appendix
\section{Exact simulation}\label{sec:exact}
Since for a prolate spheroid the length of the semi-major axis equals the radius of the smallest circumscribed ball the distribution of this radius is directly given which makes the following exact simulation method quite efficient. An implementation for spheroids is also part of the contributed \textsf{R} package \textsf{unfoldr}, see \citet{ref:Baaske2017}.

Denoting by $B_r(x)$ a ball in $\mathbb{R}^3$ with centre $x\in\mathbb{R}^3$ and radius $r\in[0,\infty)$ let $[W]=\{(x,r)\in\mathbb{R}^3\times[0,\infty):\,B_r(x)\cap W\neq\varnothing\}$ for some convex compact set $W\subset\mathbb{R}^3$. Let $\Psi$ be a marked point process on $\mathbb{R}^3\times[0,\infty)$ representing a stationary Poisson ball process \citep[Sect. 4.1]{ref:SchneiderWeil2008} with intensity $\lambda$ and p.d.f. $f_R$ (with finite third moment) of the balls' radii. Then the mean number of balls from $\Psi$ hitting $W$ equals
\begin{align*}
\Lambda_{\Psi}([W])&=\int_{\mathbb{R}^3}\int_0^{\infty}\!\!\!\lambda\,f_R(r)\,\mathbf{1}_{[W]}(x,r)\,\operatorname{d}\!r\operatorname{d}\!x=\int_{\mathbb{R}^3}\int_0^{\infty}\!\!\!\lambda\,f_R(r)\,\mathbf{1}_{W\oplus B_r(o)}(x)\,\operatorname{d}\!r\operatorname{d}\!x\\
&=\int_0^{\infty}\!\!\!\!\lambda\,f_R(r)\,V(W\oplus B_r(o))\,\operatorname{d}\!r=\int_0^{\infty}\!\!\!\!\lambda\,f_R(r)\,\sum_{k=0}^3a_kr^k\,\operatorname{d}\!r
=\lambda\sum_{k=0}^3a_k\mathsf{E}[R^k],
\end{align*}
where $W\oplus B_r(o)=\{x+y:\,x\in W,y\in B_r(o)\}$ and $R$ denotes the random ball radius, and where we have used first Fubini's theorem and then Steiner's formula; the latter implying $a_0=V(W)$, $a_1=S(W)$ (the surface area of $W$), $a_2=M(W)$ (the integral of mean curvature of $W$) and $a_3=4\pi/3$, see \citet[p. 15f.]{ref:CSKM2013} and \citet[p. 600ff.]{ref:SchneiderWeil2008}.

Hence, $\Psi([W])$ follows a Poisson distribution with mean $\lambda\sum_{k=0}^3a_k\mathsf{E}[R^k]$. Given $\Psi([W])=n$ the corresponding $n$ points $(x_1,r_1),\ldots,(x_n,r_n)$ from $\Psi$ are i.i.d. according to $\Lambda_{\Psi}(\cdot)/\Lambda_{\Psi}([W])$, i.\,e. with joint p.d.f.
\begin{equation*}
f^{[W]}(x,r)=\frac{\lambda\,f_R(r)\,\mathbf{1}_{W\oplus B_r(o)}(x)}{\Lambda_{\Psi}([W])},
\end{equation*}
see also \citet[Prop. 13.2.1]{ref:Lantuejoul2002} and \citet{ref:Lantuejoul2013}.
This implies that the radii $r_1,\ldots,r_n$ are distributed according to the p.d.f.
\begin{equation*}
f^{[W]}(r)=\int_{\mathbb{R}^3}f^{[W]}(x,r)\,\mathrm{d}x=\frac{\lambda\,f_R(r)\,V(W\oplus B_r(o))}{\Lambda_{\Psi}([W])}=\sum_{k=0}^3\frac{a_kr^kf_R(r)}{\sum_{j=0}^3a_j\mathsf{E}[R^j]}
\end{equation*}
and, given the radius $r$, the centre $x$ is uniformly distributed on $W\oplus B_r(o)$,
\begin{equation*}
f^{[W]}(x|r)=\frac{f^{[W]}(x,r)}{f^{[W]}(r)}=\frac{\mathbf{1}_{W\oplus B_r(o)}(x)}{V(W\oplus B_r(o))}.
\end{equation*}
Hence, for the exact simulation of a stationary Poisson ball process w.r.t. the window $W$, the following algorithm \citep[Algorithm 1 in][]{ref:Lantuejoul2013} is suitable:\\[0.2em]
\begin{alg}\label{alg:genPoi}
\quad
\begin{itemize}
\item[1.] Generate $n$ according to the Poisson distribution with mean $\Lambda_{\Psi}([W])$.
\item[2.] Generate $r_1,\ldots,r_n$ i.i.d. according to $f^{[W]}(r)$.
\item[3.] Generate $x_1,\ldots,x_n$ independently and, each conditionally on $r_i$, uniformly on $W\oplus B_{r_i}(o)$, $i=1,\ldots,n$. 
\end{itemize}
\end{alg}
\bigskip

In the most crucial step 2 in Algorithm \ref{alg:genPoi} the distribution is a mixture, i.\,e. in terms of the respective c.d.f. $F^{[W]}(r)$ we have 
\begin{equation*}
F^{[W]}(r)=\sum_{k=0}^3\frac{a_k\mathsf{E}[R^k;R\leq r]}{\sum_{j=0}^3a_j\mathsf{E}[R^j]}=\sum_{k=0}^3\frac{a_k\mathsf{E}[R^k]}{\sum_{j=0}^3a_j\mathsf{E}[R^j]}\,\frac{\mathsf{E}[R^k;R\leq r]}{\mathsf{E}[R^k]}
\end{equation*}
with mixing weights
\begin{equation*}
p_k=\frac{a_k\mathsf{E}[R^k]}{\sum_{j=0}^3a_j\mathsf{E}[R^j]}.
\end{equation*}
Since the random ball radius $R$ has a log-normal distribution $\mathsf{logN}(\mu,\sigma^2)$, i.\,e.
\begin{equation*}
f_R(r)=\frac{1}{\sqrt{2\pi}\sigma\,r}\exp\left(-\frac{(\log(r)-\mu)^2}{2\sigma^2}\right),
\end{equation*} 
where $\mu\in\mathbb{R}$ and $\sigma>0$, we have -- as a somewhat nice fact -- that for all $k=0,1,\ldots$
\begin{equation*}
\frac{\mathsf{E}[R^k;R\leq r]}{\mathsf{E}[R^k]}=\Phi\left(\frac{\log(r)-(\mu+k\sigma^2)}{\sigma}\right)
\end{equation*}
(where $\Phi$ denotes the c.d.f. of the standard normal distribution) is the c.d.f. of a log-normal distribution $\mathsf{logN}(\mu+k\sigma^2,\sigma^2)$. Likewise we have $\mathsf{E}[R^k]=\exp(k\mu+k^2\sigma^2/2)$, $k=0,1,2,3$. Hence, in order to generate random numbers according to $f^{[W]}(r)$ we can apply the following algorithm \citep[similar to Algorithm 2 in][]{ref:Lantuejoul2013}:\\[0.2em]
\begin{alg}
\quad
\begin{itemize}
\item[1.] Generate a random integer $k$ from $\{0,1,2,3\}$ according to the probabilities
\begin{equation*}
p_k=\frac{a_k\exp(k\mu+k^2\sigma^2/2)}{\sum_{j=0}^3a_j\exp(j\mu+j^2\sigma^2/2)}.
\end{equation*}
\item[2.] Deliver a random number from a $\mathsf{logN}(\mu+k\sigma^2,\sigma^2)$-distribution.
\end{itemize}
\end{alg}
\bigskip

Once all circumscribed balls hitting $W$ are generated, subsequently the shapes and orientations (see \ref{sec:orient}) of the prolate spheroids can be generated. In case one is interested only in a configuration of spheroids hitting $W$ then possibly a few non-hitting spheroids have to be deleted from the configuration as the final step.



\section{Simulation of orientations}\label{sec:orient}
From (\ref{eqn:pbeta}) we have 
\begin{equation}
h_{\beta}(\vartheta)=\frac{1}{2}\frac{\beta\sin\vartheta}{(1+(\beta^2-1)\cos^2\vartheta)^{\frac{3}{2}}},\quad\vartheta\in[0,\pi),
\end{equation}
and
\begin{equation}
H_{\beta}(\vartheta)=\frac{1}{2}\left(1-\frac{\beta\cos\vartheta}{(1+(\beta^2-1)\cos^2\vartheta)^{\frac{1}{2}}}\right),\quad\vartheta\in[0,\pi),
\end{equation}
for, respectively, the p.d.f. and the c.d.f. of the polar angle $\vartheta$. This in turn leads to the quantile function
\begin{equation}
H_{\beta}^{-1}(q)=\arccos\left(\frac{1-2q}{\sqrt{\beta^2-(1-2q)^2(\beta^2-1)}}\right)
\end{equation}
which can be used to generate random numbers according to $H_{\beta}$ with the inversion method.

\section{Bootstrap standard errors}\label{sec:bootstrap}
In Tables \ref{tab:sderr1}--\ref{tab:sderr5} the bootstrap standard errors of the estimated root mean squared errors given in Tables \ref{tab:rmse1}--\ref{tab:rmse5} are provided.
\begin{table}
\centering
\caption{Bootstrap standard errors of the estimated root mean squared errors in Table \ref{tab:rmse1}.}
\label{tab:sderr1}
\begin{tabular}{l|rrrrrr}
Method&rmse$(\hat{\mu}_1)$&rmse$(\hat{\mu}_2)$&rmse$(\hat{\sigma}_1)$&rmse$(\hat{\sigma}_2)$&rmse$(\hat{\varrho})$&rmse$(\hat{\beta}$)\\
\hline
UMLE$_1$&0.0040&0.0053&0.0025&0.0042&0.0127&0.0038\\
ULME$_2$&0.0042&0.0036&0.0035&0.0022&0.0124&0.0044\\
ULME$_3$&0.0041&0.0035&0.0043&0.0026&0.0090&0.0046\\
ULME$_4$&0.0041&0.0029&0.0046&0.0024&0.0093&0.0044\\
ULME$_5$&0.0039&0.0031&0.0045&0.0028&0.007&0.0045\\
QLE&0.0019&0.0028&0.0020&0.0028&0.0143&0.0031\\
BINMLE$_1$&0.0008&0.0032&0.0009&0.0080&0.0068&0.0022\\
BINMLE$_2$&0.0007&0.0009&0.0008&0.0006&0.0019&0.0029\\
BINMLE$_3$&0.0009&0.0006&0.0006&0.0006&0.0027&0.0022\\
BINMLE$_4$&0.0008&0.0008&0.0006&0.0007&0.0021&0.0023\\
BINMLE$_5$&0.0008&0.0007&0.0005&0.0006&0.0022&0.0027\\
MLE3D&0.0008&0.0007&0.0006&0.0004&0.0032&0.0031
\end{tabular}
\end{table}
\begin{table}
\centering
\caption{Bootstrap standard errors of the estimated root mean squared errors in Table \ref{tab:rmse2}.}
\label{tab:sderr2}
\begin{tabular}{l|rrrrrr}
Method&rmse$(\hat{\mu}_1)$&rmse$(\hat{\mu}_2)$&rmse$(\hat{\sigma}_1)$&rmse$(\hat{\sigma}_2)$&rmse$(\hat{\varrho})$&rmse$(\hat{\beta}$)\\
\hline
UMLE$_1$&0.0036&0.0086&0.0024&0.0041&0.0201&0.0189\\
ULME$_2$&0.0037&0.0033&0.0036&0.0021&0.0104&0.0181\\
ULME$_3$&0.0044&0.0022&0.0048&0.0025&0.0093&0.0197\\
ULME$_4$&0.0047&0.0041&0.0063&0.0027&0.0079&0.0214\\
ULME$_5$&0.0045&0.0039&0.0056&0.0028&0.0071&0.0223\\
QLE&0.0027&0.0043&0.0020&0.0032&0.0242&0.0340\\
BINMLE$_1$&0.0008&0.0029&0.0009&0.0072&0.0056&0.0519\\
BINMLE$_2$&0.0008&0.0010&0.0006&0.0006&0.0024&0.0420\\
BINMLE$_3$&0.0007&0.0007&0.0006&0.0005&0.0021&0.0465\\
BINMLE$_4$&0.0007&0.0008&0.0005&0.0006&0.0024&0.0317\\
BINMLE$_5$&0.0011&0.0007&0.0005&0.0005&0.0024&0.0241\\
MLE3D&0.0006&0.0006&0.0005&0.0004&0.0020&0.0266
\end{tabular}
\end{table}
\begin{table}
\centering
\caption{Bootstrap standard errors of the estimated root mean squared errors in Table \ref{tab:rmse3}.}
\label{tab:sderr3}
\begin{tabular}{l|rrrrrr}
Method&rmse$(\hat{\mu}_1)$&rmse$(\hat{\mu}_2)$&rmse$(\hat{\sigma}_1)$&rmse$(\hat{\sigma}_2)$&rmse$(\hat{\varrho})$&rmse$(\hat{\beta}$)\\
\hline
UMLE$_1$&0.0039&0.0085&0.0027&0.0053&0.0148&0.0025\\
ULME$_2$&0.0037&0.0034&0.0036&0.0019&0.0096&0.0022\\
ULME$_3$&0.0034&0.0031&0.0045&0.0022&0.0085&0.0024\\
ULME$_4$&0.0039&0.0031&0.0047&0.0024&0.0081&0.0024\\
ULME$_5$&0.0042&0.0029&0.0060&0.0027&0.0069&0.0023\\
QLE&0.0021&0.0021&0.0017&0.0022&0.0134&0.0018\\
BINMLE$_1$&0.0011&0.0035&0.0009&0.0086&0.0076&0.0012\\
BINMLE$_2$&0.0009&0.0008&0.0008&0.0007&0.0031&0.0014\\
BINMLE$_3$&0.0007&0.0007&0.0006&0.0005&0.0026&0.0014\\
BINMLE$_4$&0.0009&0.0010&0.0009&0.0005&0.0027&0.0014\\
BINMLE$_5$&0.0007&0.0008&0.0006&0.0005&0.0022&0.0011\\
MLE3D&0.0008&0.0007&0.0006&0.0005&0.0027&0.0010
\end{tabular}
\end{table}
\begin{table}
\centering
\caption{Bootstrap standard errors of the estimated root mean squared errors in Table \ref{tab:rmse4}.}
\label{tab:sderr4}
\begin{tabular}{l|rrrrrr}
Method&rmse$(\hat{\mu}_1)$&rmse$(\hat{\mu}_2)$&rmse$(\hat{\sigma}_1)$&rmse$(\hat{\sigma}_2)$&rmse$(\hat{\varrho})$&rmse$(\hat{\beta}$)\\
\hline
UMLE$_1$&0.0042&0.0055&0.0028&0.0044&0.0141&0.0039\\
ULME$_2$&0.0036&0.0044&0.0037&0.0019&0.0158&0.0042\\
ULME$_3$&0.0036&0.0044&0.0040&0.0031&0.0115&0.0042\\
ULME$_4$&0.0038&0.0024&0.0042&0.0024&0.0109&0.0039\\
ULME$_5$&0.0034&0.0037&0.0043&0.0027&0.0093&0.0039\\
QLE&0.0023&0.0038&0.0023&0.0035&0.0201&0.0029
\end{tabular}
\end{table}
\begin{table}
\centering
\caption{Bootstrap standard errors of the estimated root mean squared errors in Table \ref{tab:rmse5}.}
\label{tab:sderr5}
\begin{tabular}{l|rrrrrr}
Method&rmse$(\hat{\mu}_1)$&rmse$(\hat{\mu}_2)$&rmse$(\hat{\sigma}_1)$&rmse$(\hat{\sigma}_2)$&rmse$(\hat{\varrho})$&rmse$(\hat{\beta}$)\\
\hline
UMLE$_1$&0.0041&0.0058&0.0029&0.0037&0.0114&0.0043\\
ULME$_2$&0.0043&0.0055&0.0033&0.0019&0.0199&0.0048\\
ULME$_3$&0.0040&0.0027&0.0042&0.0024&0.0134&0.0044\\
ULME$_4$&0.0038&0.0035&0.0044&0.0023&0.0123&0.0046\\
ULME$_5$&0.0034&0.0037&0.0043&0.0027&0.0109&0.0044\\
QLE&0.0058&0.0043&0.0135&0.0285&0.0420&0.0058
\end{tabular}
\end{table}
%


\bibliographystyle{elsarticle-harv}     
\bibliography{spheroidest} 
\end{document}